\documentclass[nofootinbib,prd,aps,superscriptaddress,preprintnumbers,preprint]{revtex4}
\usepackage{graphicx}
\usepackage{epstopdf}
\usepackage{amsfonts}
\usepackage{amsmath}
\usepackage{ulem}
\usepackage{bm}
\usepackage{epsfig}
\usepackage{graphics}
\usepackage{xspace}
\usepackage[usenames]{color}
\usepackage{subfigure}
\usepackage{feynmf}
\usepackage{bbold}
\def\bea{\begin{eqnarray}}
\def\eea{\end{eqnarray}}
\def\ba{\begin{eqnarray}}
\def\ea{\end{eqnarray}}
\def\be{\begin{equation}}
\def\ee{\end{equation}}
\def\beq{\begin{equation}}
\def\eeq{\end{equation}}

\newcommand{\lsim}{\mathrel{\rlap{\lower4pt\hbox{\hskip1pt$\sim$}}
    \raise1pt\hbox{$<$}}}         
\newcommand{\gsim}{\mathrel{\rlap{\lower4pt\hbox{\hskip1pt$\sim$}}
    \raise1pt\hbox{$>$}}}         

\newcommand{\leftrightarrowraised}{\mathrel{\rlap{\lower-0pt\hbox{\hskip1pt$\partial$}}
    \raise6 pt\hbox{$\leftrightarrow$}}}

\DeclareMathOperator{\tr}{tr}

\def\1{ \mathbb{1}_\epsilon}

\def\tr{\text{Tr}\,}

\begin{document}

\title{A consistent picture for large penguins in $D\to \pi^+ \pi^- \,,\,K^+ K^- $}

\def\Cincy{Department of Physics, University of Cincinnati, Cincinnati, OH 45221,USA}
\author{Joachim Brod}
\email[Electronic address:]{brodjm@ucmail.uc.edu} 
\affiliation{\Cincy}
\author{Yuval Grossman}
\email[Electronic address:]{yg73@cornell.edu} 
\affiliation{Department of Physics, LEPP, Cornell University, Ithaca,
  NY 14853, USA}
\author{Alexander L. Kagan}
\email[Electronic address:]{kaganalexander@gmail.com} 
\affiliation{\Cincy}
\author{Jure Zupan}
\email[Electronic address:]{jure.zupan@cern.ch}

\affiliation{\Cincy}

\begin{abstract}
A long-standing puzzle in charm physics is the large difference
between the $D^0\to K^+K^-$ and $D^0\to \pi^+ \pi^-$ decay
rates. Recently, the LHCb and CDF collaborations reported a
surprisingly large difference between the direct CP asymmetries,
$\Delta {\cal A}_{CP}$, in these two modes.  We show that the two
puzzles are naturally related in the Standard Model via $s$- and
$d$-quark ``penguin contractions". Their sum gives rise to $\Delta
{\cal A}_{CP}$, while their difference contributes to the two
branching ratios with opposite sign. Assuming nominal SU(3) breaking,
a U-spin fit to the $D^0 \to K^+ \pi^-,\pi^+ K^- \,,\pi^+ \pi^- \,,
K^+ K^-$ decay rates yields large penguin contractions that naturally
explain $\Delta {\cal A}^{}_{CP}$.  Expectations for the individual CP
asymmetries are also discussed.

\end{abstract}

\maketitle
\newpage
\section{Introduction}
There are several surprising experimental facts in $D^0$ decays to
pairs of charged pseudoscalars.  The first one is the long-standing
puzzle of the large rate difference between the singly
Cabibbo-suppressed (SCS) decays, $D^0 \to K^+K^- $ and $D^0 \to
\pi^+\pi^- $. 
The two amplitudes would be equal in 
the U-spin symmetric limit, whereas the measured rates yield
\beq
\left|{ {\cal A}(  D^0 \to K^+K^-) \over  {\cal A}(  D^0  \to \pi^+\pi^-)}\right| -1  =(0.82\pm 0.02), \label{Exp-amplitudes}
\eeq
for their CP averaged magnitudes, after accounting for phase space.\footnote{We define
$|{\cal A}(D^0\to PP)|\equiv [\,{\overline \Gamma(D^0 \to PP)} \,8 \pi
m_D^2 /p_c \,]^{1/2}/(1\,{\rm keV})$,
where $\overline \Gamma$ is the CP averaged decay rate and $p_c$ is the center-of-mass momentum of the final state mesons.}
This has led to speculation that U-spin breaking in SCS $D$ decays is
${\mathcal O}(1)$ \cite{Savage:1991wu,Hinchliffe:1995hz,Ryd:2009uf,Cheng:2010ry,Pirtskhalava:2011va,Cheng:2012wr,Bhattacharya:2012ah,Feldmann:2012js}, rather than of the nominal size characterized by 
\beq \epsilon_{U}\sim (f_K/f_\pi -1)\sim {\mathcal O}(0.2)\,.\label{nominal}\eeq 
However, such a conclusion is premature, as indicated by the following 
interesting experimental observations:
\begin{enumerate}
\item
The CP averaged magnitudes for the Cabibbo-favored (CF) $D^0 \to K^- \pi^+  $ and doubly Cabibbo-suppressed (DCS) 
$D^0 \to K^+ \pi^- $ amplitudes satisfy
\beq \label{kpi-exp}
\left| { V_{cs} V_{ud}\over   V_{cd} V_{us}} \,\,{{\cal A} (D^0 \to K^+ \pi^-  ) \over  {\cal A} (D^0 \to K^- \pi^+  ) }\right| -1 
= (15.1\pm 2.8)\% \,.
\eeq
\item
The CP averaged amplitudes satisfy the experimental ``sum-rule" relation
(a similar sum rule has been discussed in \cite{Falk:2001hx})
\beq\label{sumrule-exp}
\Sigma_{\text{sum-rule}} = { \left|{\cal A}(  D^0 \to K^+K^-)\,/\,V_{cs} V_{us} \right| +  \left|{\cal A}(  D^0 \to \pi^+ \pi ^-)\,/\,V_{cd} V_{ud} \right|
\over \left| {\cal A} (D^0 \to K^+ \pi^-  )\, /\, V_{cd} V_{us} \right| + 
\left| {\cal A} (D^0 \to K^- \pi^+  )\, /\, V_{cs} V_{ud} \right| } -1 =
(4.0\pm 1.6)\%.
\eeq
\end{enumerate}
The expressions in \eqref{kpi-exp} and \eqref{sumrule-exp} would
vanish in the U-spin limit. Thus, the fact that they are small
experimentally suggests that U-spin is a good approximate symmetry in
these decays.  The alternative is that U-spin breaking is ${\mathcal
O}(1)$ in the SCS decays under consideration, and nominal in the
CF/DCS ones.  However, such a hierarchy of U-spin breakings would be
left unexplained.

Another interesting result is the surprisingly large time-integrated
CP asymmetry difference, 
$\Delta {\cal A}_{CP} \equiv {\cal A}_{CP} (D^0 \to K^+ K^- ) - {\cal
  A}_{CP} (D^0 \to \pi^+ \pi^- )$, 
recently measured by the LHCb and CDF
collaborations~\cite{Aaij:2011in,CDF-talk}.  Inclusion of the Babar
and Belle measurements of the individual $K^+ K^-$ and $ \pi^+ \pi^- $ 
time-integrated CP asymmetries \cite{Aubert:2007if,Staric:2008rx} and
the indirect CP asymmetry $A_\Gamma  $
\cite{Aubert:2007en,Staric:2007dt} yields the world average for the
direct CP asymmetry difference \cite{CDF-talk} 
\beq
\Delta {\cal A}^{\rm dir}_{CP}\equiv {\cal A}^{\rm dir}_{CP}(D\to K^+K^-) -
{\cal A}^{\rm dir}_{CP}(D\to \pi^+\pi^-) = (-0.67 \pm 0.16)\,\%.  \label{DeltaACP}
\eeq

In the Standard Model (SM), the ratio of penguin-to-tree amplitudes is
naively of $\mathcal O([V_{cb} V_{ub}/V_{cs} V_{us}] \alpha_s/\pi
)\sim 10^{-4}$, yielding $\Delta {\cal A}^{\rm dir}_{CP}< 0.1\%$.
This expectation is based on estimates of the ``short-distance"
penguins with $b$-quarks in the loops.  While $\Delta A^{\rm
  dir}_{CP}$ could be a signal of new physics \cite{Grossman:2006jg,
  Isidori:2011qw, Wang:2011uu, Hochberg:2011ru, Chang:2012gn,
  Giudice:2012qq, Altmannshofer:2012ur, Chen:2012am, Feldmann:2012js},
it was argued long ago that long-distance effects could conceivably
give large direct CP violation (CPV) due to hadronic enhancement of
penguin amplitudes~\cite{Golden:1989qx}.  Small CPV was subsequently
predicted in~\cite{Buccella:1994nf} using a model for final-state
interactions. Making use of lessons learned from the $D\to PP$ data
that has since become available, it was estimated
in~\cite{Brod:2011re} (see also \cite{kaganFPCP}) that formally
power-suppressed long-distance $s$- and $d$-quark ``penguin
contractions" can yield penguin-to-tree ratios of ${\mathcal O}(0.1
\%)$, thus potentially explaining the observed $\Delta {\cal A}^{\rm
  dir}_{\rm CP}$.  More recent works argue that a SM explanation is
either marginal \cite{Franco:2012ck}, or not possible
\cite{Li:2012cf}.

In this paper we show that the possibility of a large penguin
amplitude in the SM acquires further support from the experimental
data.  A consistent picture emerges in which
\begin{enumerate} 
\item U-spin breaking is nominal.  This helps explaining
  \eqref{kpi-exp} and \eqref{sumrule-exp} with minimal tuning of
  strong phases.
\item
The U-spin invariant sum of the $s$- and $d$-quark penguin contractions
enhances the penguin amplitude, thus explaining $\,\Delta {\cal
A}^{\rm dir}_{CP}\,$ in \eqref{DeltaACP}.
\item
The difference between the $s$- and $d$-quark penguin contractions,
which we refer to as the ``broken penguin" (with respect to U-spin),
explains the difference in the decay amplitudes in
\eqref{Exp-amplitudes}.
\end{enumerate}

The last point above requires a broken penguin amplitude that is of
the same order as the tree amplitude (with a $\sim 50\%$ smaller value
preferred), yielding substantial interference between the two (see
also~\cite{Bhattacharya:2012ah}). In turn, nominal U-spin breaking
implies that the penguin amplitude is enhanced relative to the tree
amplitude by ${\mathcal O}(1/\epsilon_U )$ (with $\sim 0.5/\epsilon_U$
preferred). This is in the favored range to explain the observed
$\Delta {\cal A}^{\rm dir}_{CP}$.

The situation we describe in this paper resembles the one which arises
in kaon decays: the apparently large isospin breaking in $K \to
\pi\pi$~\cite{Cirigliano:2009rr} results from a combination of nominal
isospin breaking and the ``$\Delta I=1/2$ rule'', i.e., the
enhancement of the $A_{0}$ amplitude relative to $A_{2}$. Here we
suggest that the apparently large U-spin breaking in $D^0 \to K^+ K^-
, \pi^+ \pi^-$ decays is a consequence of both nominal U-spin breaking
and an enhancement of the $\Delta U=0$ penguin matrix elements
relative to the tree amplitude.

The paper is organized as follows.  In Section~\ref{sec:pen} we
discuss the relevant $\Delta C=1$ effective Hamiltonian, explain our
counting in $\epsilon_U$ for the operator matrix elements, and give
the U-spin decomposition of the decay amplitudes.
Note that our counting is modified in order to take explicit account
of the penguin contractions, thus allowing for their enhancement.
Implications of the measured CP averaged decay rates for the penguin
contractions are then studied.  In Section~\ref{sec:CP} we incorporate
the CP violation data.  We conclude in Section~\ref{sec:concl}.  A
derivation of the U-spin decomposition together with U-spin breaking
is given in
Appendix~\ref{Formal-U-spin-app}. Appendix~\ref{App:diagrammatics}
contains the equivalent ``diagrammatic" decomposition of the decay
amplitudes.  We also provide a translation between the diagrammatic
and the U-spin reduced matrix elements.


\section{Enhanced penguins and decay rates}\label{sec:pen}
The decays 
\beq
\bar D^0\to K^+\pi^-,\qquad
\bar D^0\to K^-\pi^+,\qquad
\bar D^0\to \pi^+\pi^-,\qquad
\bar D^0\to K^+K^-
\eeq
are related through U-spin. We begin with a discussion of their U-spin
decomposition.  Using Cabibbo-Kobayashi-Maskawa (CKM) unitarity we can write the Hamiltonian governing
SCS decays as
\begin{equation}
\begin{split}
H_{\rm eff}^{\rm SCS} = \frac{G_F}{\sqrt{2}} & \left\{
\left(V_{cs} V_{us}^*- V_{cd} V_{ud}^*\right)  \sum_{i=1,2} C_i \left(Q_i^{\bar s s} - Q_i^{\bar dd} \right)/2 \right. \\
&\left. -V_{cb} V_{ub}^* \,\left[ \sum_{i=1,2} C_i  \left(Q_i^{\bar s s} + Q_i^{\bar d d} \right)/2  +  \sum_{i=3}^{6} 
C_i Q_i + C_{8g} Q_{8 g} \,\,\right]\right\}
+ {\rm h.c.}\,,
\label{eq:Heff}
\end{split}
\end{equation}
where 
\beq
Q_1^{\bar p p'}=(\bar p u)_{V-A}(\bar c p')_{V-A}, \qquad Q_2^{\bar p p'}=(\bar p_\alpha u_\beta)_{V-A} (\bar c_\beta p'_\alpha)_{V-A}
\eeq
 are the ``tree operators", $Q_{3,..,6}$ are the QCD penguin operators, and $Q_{8g}$ is the chromomagnetic dipole operator.
The effective Hamiltonian for Cabibbo-favored (CF)  decays contains only the tree operators, 
\begin{equation}
\begin{split}
H_{\rm eff}^{\rm CF} = \frac{G_F}{\sqrt{2}} & 
V_{cs} V_{ud}^*   \sum_{i=1,2} C_i Q_i^{\bar d s}
+ {\rm h.c.},
\label{eq:Heff:CF}
\end{split}
\end{equation}
and similarly for doubly Cabibbo suppressed (DCS) decays with the replacement $s\leftrightarrow d$. 

We first work in the limit in which U-spin is exact or nearly so, but
with the $d$ and $s$ quarks still of two distinguishable flavors. 
Our working assumption is that the penguin contraction contributions
coming from the $Q_{1,2}$ operators are enhanced compared to the
tree amplitude, defined below. The enhancement is
parametrized  by $1/\epsilon'$, where $\epsilon'\ll 1$. Thus, at a
scale $\mu\sim m_D$ we have for the ${\mathcal O}(1/\epsilon')$ U-spin
invariant matrix elements  
\begin{align}
\begin{split} \label{eq:P}
P\equiv\, & \langle K^+K^-| \sum_{i=1,2} C_i Q_i^{\bar d d} \,+ \sum_{i=3,..,8g} C_i Q_i |\bar D^0\rangle= \\
&\qquad\qquad\qquad =  \langle \pi^+\pi^-| \sum_{i=1,2} C_i Q_i^{\bar s s}\,+ \sum_{i=3,..,8g} C_i Q_i |\bar D^0\rangle \sim {\mathcal O}(1/\epsilon'),
\end{split}
\\
\begin{split}\label{TandP} 
T+P\equiv \, & \langle K^+K^-| \sum_{i=1,2} C_i Q_i^{\bar s s} \,+ \sum_{i=3,..,8g} C_i Q_i |\bar D^0\rangle=\\
& \qquad\qquad\qquad=\langle \pi^+\pi^-| \sum_{i=1,2} C_i Q_i^{\bar d d}\, + \sum_{i=3,..,8g} C_i Q_i |\bar D^0\rangle \sim {\mathcal O}(1/\epsilon').
\end{split}
\end{align}
The $P$ amplitude is a pure $\Delta U=0$ transition.  Note that the $Q_{1,2}$ operators in \eqref{eq:P} can only produce the $K^+K^-$  final state if the $\bar dd$ quark pairs are
contracted (and similarly for $\bar ss$ in the case of $\pi^+\pi^-$).
The contractions are illustrated diagrammatically in Fig.~\ref{fig:T} of Appendix \ref{App:diagrammatics}.  The contractions also have a non-perturbative field theoretic definition which employs the lattice as a UV regulator. 
This is what we mean when we refer to the penguin contractions below. In  \eqref{TandP} the non-contracted contributions of $Q_{1,2}$ are part of $T$.   

The contributions of the penguin operators $Q_{3,..,6}$, $Q_{8g}$ in
\eqref{eq:P} and \eqref{TandP} are expected to be an order of
magnitude smaller than required to explain $\Delta {\cal A}_{CP}^{\rm
  dir}$, see e.g., \cite{Brod:2011re}, and are thus ignored throughout
this work.  Note, however, that the scheme and renormalization-scale
dependence in their Wilson coefficients cancels the scheme and scale
dependence appearing in the penguin contraction matrix elements of
$Q_{1,2}$ in \eqref{eq:P} and \eqref{TandP}.  This cancelation is
understood whenever we refer to $P$ below.

The $T$ matrix element is ${\mathcal O}(1)$ in the $\epsilon'$ counting. In the U-spin limit it is given by
\beq
\begin{split}
T&=-\frac{1}{2}\Big(\langle K^+K^-| \sum_{i=1,2} C_i (Q_i^{\bar d d}-Q_i^{\bar s s})|\bar D^0\rangle - \langle \pi^+\pi^-| \sum_{i=1,2} C_i (Q_i^{\bar d d}- Q_i^{\bar s s})|\bar D^0\rangle\Big)= \\
&= \langle K^+\pi^-| \sum_{i=1,2} C_i Q_i^{\bar d s}|\bar D^0\rangle =\langle \pi^+K^-| \sum_{i=1,2} C_i Q_i^{\bar s d}|D\rangle\sim {\mathcal O}(1) \,.
\label{eq-t1}
\end{split}
\eeq
This follows from the fact that both $\langle K^+\pi^-|$, $(\langle
K^+K^-|-\langle \pi^+\pi^-|)/\sqrt2$, $\langle K^-\pi^+|$ and
$Q_i^{\bar d s}$, $(Q_i^{\bar s s}-Q_i^{\bar d d})/\sqrt2$, $Q_i^{\bar
  s d}$ are U-spin triplets. 

The sum of the two amplitudes in the first line of \eqref{eq-t1} is
U-spin breaking, giving a ``broken penguin",
\beq  
\begin{split}
 P_{\rm break}\equiv \frac{1}{2}\bigg(\langle K^+K^-| \sum_{i=1,2} C_i (Q_i^{\bar d d}-Q_i^{\bar s s})|\bar D^0\rangle + \langle \pi^+\pi^-| \sum_{i=1,2} C_i (Q_i^{\bar d d}- Q_i^{\bar s s})|\bar D^0\rangle\bigg).\label{eq:t3}
\end{split}
\eeq
The leading contribution to $P_{\rm break} $ measures the difference
between final-state interactions involving the $\bar s s$ and $\bar d
d $ contractions.  The broken penguin is parametrically of the size  
\beq
P_{\rm break}\sim \epsilon_U P\sim {\mathcal O}(\epsilon_U/\epsilon')\sim {\mathcal O}(1).
\eeq
In the last equality we have used the scaling $\epsilon'\sim
\epsilon_U$, which is satisfied by the data, as shown below. For now,
however, we keep $\epsilon'$ and $\epsilon_U$ separate.

As already stressed, our working assumption is that the  matrix
elements containing penguin contractions of $Q_{1,2}$ operators are
enhanced. We then have two sets of matrix elements, the ones that are
${\mathcal O}(1/\epsilon')$ enhanced, and the ones that are not. For
each of them there is also an expansion in the U-spin breaking
parameter $\epsilon_U$. At $n$-th
order in U-spin breaking the reduced amplitudes that are not enhanced
are of  ${\mathcal O}(\epsilon_U^n)$, while the reduced amplitudes
that contain penguin contractions are of ${\mathcal
  O}(\epsilon_U^n/\epsilon')$. For example, summarizing the above
results, we have the following scalings 
\beq
T\sim {\mathcal O}(1), \qquad P\sim {\mathcal O}(1/\epsilon'), \qquad P_{\rm break} \sim {\mathcal O}(\epsilon_U/\epsilon').
\eeq

The expressions \eqref{eq-t1} and \eqref{eq:t3} are valid to ${\mathcal
O}(1,\epsilon_U/\epsilon')$.  At ${\mathcal O}(\epsilon_U,
\epsilon_U^2/\epsilon')$ the sum of matrix elements in \eqref{eq:t3}
also receives a contribution due to U-spin breaking in $T$, changing
the l.h.s. from $P_{\rm break}\to P_{\rm
break}(1-\tfrac{1}{2}\epsilon_{sd}^{(2)})+\tfrac{1}{2}\epsilon_{T1}
T$.  In principle, the two contributions -- the $P_{\rm break} \sim
\epsilon_U P$ term and the $\epsilon_U T$ term -- could be
separated, if necessary.  They correspond to two different topological
amplitudes, with the $\bar q q$ fermion fields in the $Q_{1,2}^{\bar q
q}$ operators either contracted or not.  In particular, the matrix
elements defined in this way could be calculated on the lattice in the
(probably not so near) future \cite{Sharpe-private}.
 
The decay amplitudes are derived in Appendices~\ref{Formal-U-spin-app}
and~\ref{App:diagrammatics}. At order ${\mathcal O}(\epsilon_U,
\epsilon_U^2/\epsilon')$, and using the notations of
Appendix~\ref{App:diagrammatics}, they read
 \begin{align}
\label{diagrammatic-begin}
A(\bar D^0\to K^+\pi^-)=&V_{cs} V_{ud}^*T(1-\tfrac{1}{2}\epsilon_{T2}),
\\
\begin{split}
A(\bar D^0\to \pi^+\pi^-)=&-\tfrac{1}{2}\left(V_{cs} V_{us}^* -V_{cd} V_{ud}^*\right)\big(T(1+\tfrac{1}{2}\epsilon_{T1}) +P_{\rm break}(1- \tfrac{1}{2}\epsilon_{sd}^{(2)})\big) \label{pipidiagrammatic}\\
&-V_{cb}^*V_{ub} \left(P(1- \tfrac{1}{2}\epsilon_P)+\tfrac{1}{2}T\right),
\end{split}
\\
\begin{split}
A(\bar D^0\to K^+K^-)=&\tfrac{1}{2}\left(V_{cs} V_{us}^* -V_{cd} V_{ud}^*\right)\big(T(1-\tfrac{1}{2}\epsilon_{T1} ) -P_{\rm break}(1+ \tfrac{1}{2}\epsilon_{sd}^{(2)})\big) \label{KKdiagrammatic}\\
&-V_{cb}^*V_{ub} \left(P(1+ \tfrac{1}{2}\epsilon_P)+\tfrac{1}{2}T\right),
\end{split}
\\
A(\bar D^0\to \pi^+K^-)=&V_{cd} V_{us}^*T(1+\tfrac{1}{2}\epsilon_{T2} ).
\label{diagrammatic-end}
\end{align}
The coefficients
multiplying the U-spin breaking parameters $\epsilon_i$ are chosen
such that typically $\epsilon_i\sim {\mathcal O}(\epsilon_U)$. 
For simplicity the $V^*_{cb} \,V_{ub}$ suppressed terms are only given
to order ${\mathcal O}(1,\epsilon_U/\epsilon')$, i.e. to first
subleading order in the expansion. The scaling of the different terms
in the $V^*_{cb} V_{ub}$ suppressed amplitudes is 
\beq \label{scalings1}
P\sim {\mathcal O}(1/\epsilon'), \qquad P \epsilon_P \sim {\mathcal O}(\epsilon_U/\epsilon'),\qquad T\sim {\mathcal O}(1).
\eeq
The ``tree" parts of the amplitude are also given to the first subleading order in the expansion, which in this case is ${\mathcal O}(\epsilon_U, \epsilon_U^2/\epsilon')$, with
\beq \label{scalings2}
T\sim  {\mathcal O}(1), \qquad P_{\rm break}\sim  {\mathcal O}(\epsilon_U/\epsilon'), \qquad T\epsilon_{T1}, T\epsilon_{T2}\sim  {\mathcal O}(\epsilon_U),\qquad P_{\rm break}\epsilon_{sd}^{(2)}\sim {\mathcal O}(\epsilon_U^2/\epsilon') .
\eeq

\begin{figure}
\includegraphics[scale=0.33]{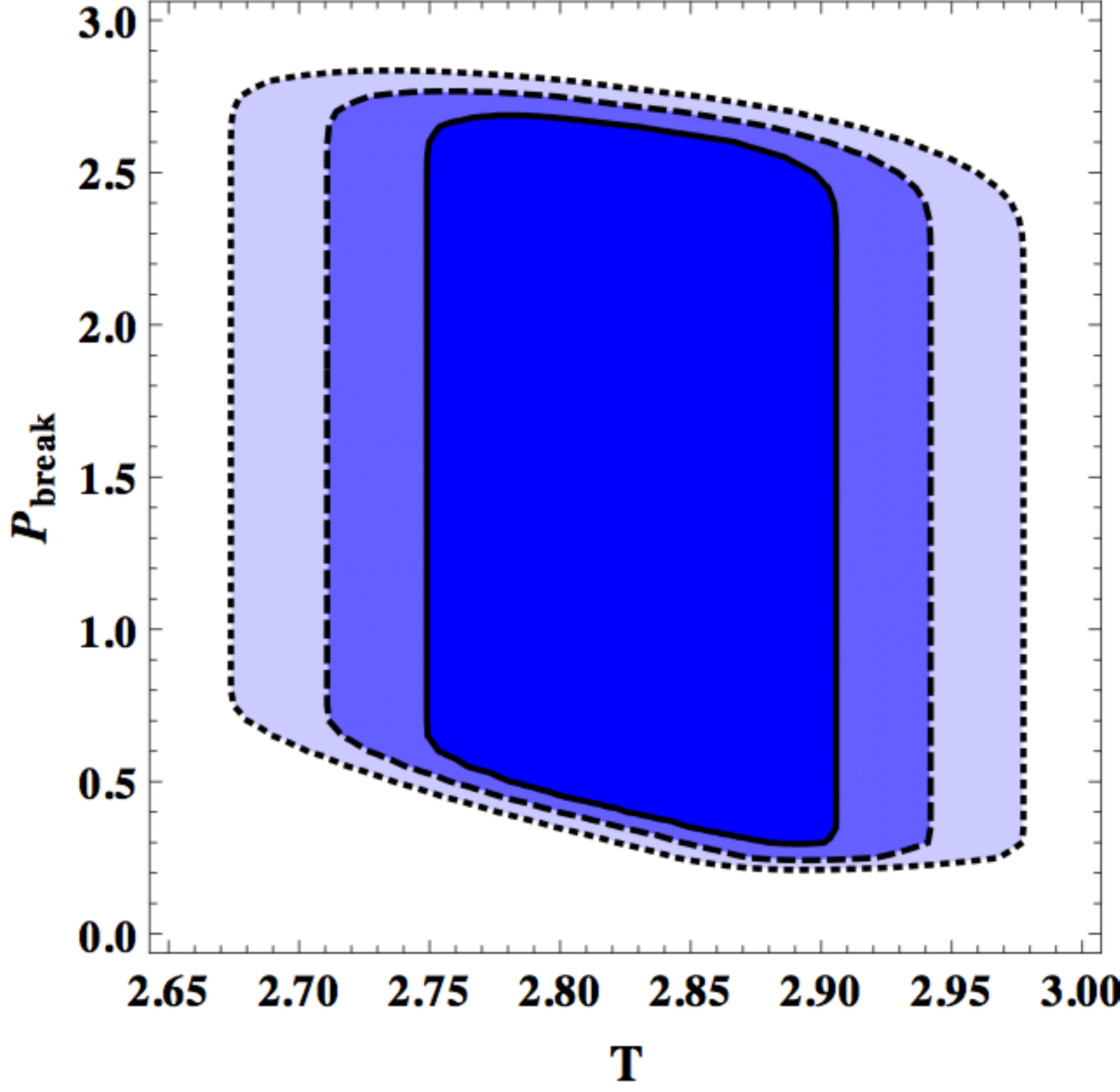}~~~~~~~~
\includegraphics[scale=0.33]{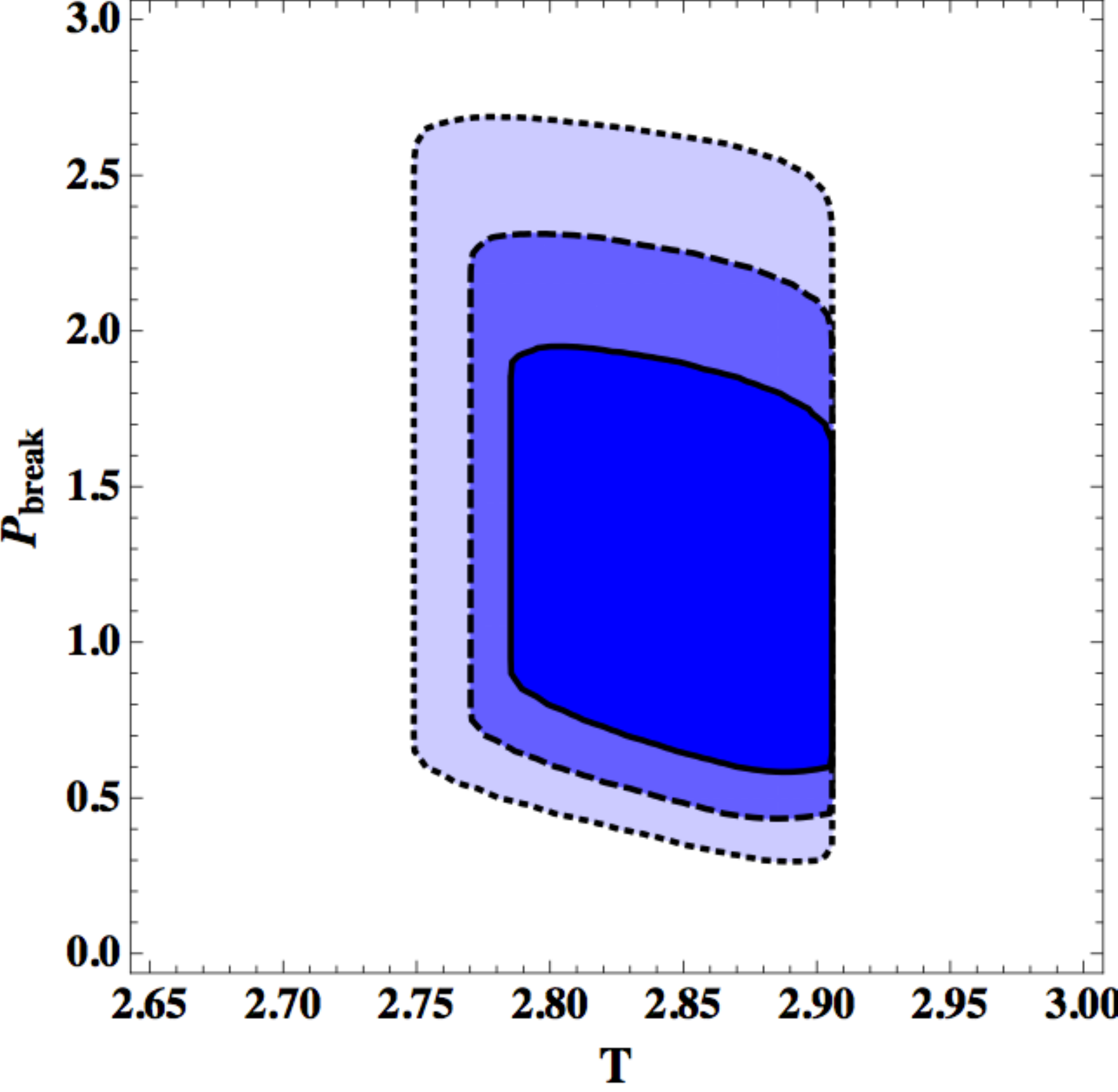}
	\caption{Constraints on $P_{\rm
            break}$ vs. $T$ from the fit to the branching ratios. Left: the contours denote regions allowed at
          $1\sigma$ (solid), $2\sigma$ (dashed), $3\sigma$
          (dotted). Right: the constraints at $1\sigma$ but with $\epsilon_i$  varied in the ranges  $\epsilon_i \in [0,0.2]$ (solid), $\epsilon_i \in [0,0.3]$ (dashed), $\epsilon_i \in [0,0.4]$ (dotted).}
	\label{FigPbT}
\end{figure}

We are now ready to check how well the implicitly assumed scaling
$\epsilon'\sim \epsilon_U$ compares with the data. In
Fig.~\ref{FigPbT} we display the result of a fit of
\eqref{diagrammatic-begin}-\eqref{diagrammatic-end} to the measured
$\bar D^0 \to K^+K^-, \pi^+\pi^-, K^\pm \pi^\mp$ branching ratios
\cite{Asner:2010qj}. 
Here we can safely neglect the $V_{cb}^*V_{ub}$-suppressed contributions. 
There are three real parameters which are floated in the fit -- the
magnitudes of $T$ and $P_{\rm break}$ and their relative strong
phase. The U-spin breaking parameters $\epsilon_{T1}$, $\epsilon_{T2}$
and $\epsilon_{sd}^{(2)}$ are varied in the constrained range
$\epsilon_i \in [0,0.4]$ for Fig.~\ref{FigPbT} (left), and in three
ranges, $\epsilon_i \in [0,0.2]$, $\epsilon_i \in [0,0.3]$, and
$\epsilon_i \in [0,0.4]$ for Fig.~\ref{FigPbT} (right).  The branching
ratios can be fit perfectly, so that the $\chi^2 $ global minimum is
$\chi_{\rm min}^2 =0$.  The fit shows that
\beq P_{\rm break}\sim T\,, \eeq 
in accord with our $\epsilon'\sim \epsilon_U$ counting.  This
is a direct consequence of the large difference between the 
$D\to K^+K^-$ and $D\to \pi^+\pi^-$ decay rates.

\section{Enhanced penguins and CP violation}\label{sec:CP}

The time-integrated CP asymmetry for SCS $D^0$ decays to a final CP eigenstate $f$ is defined as 
\beq
{\cal A}_{CP}(D\to f)  \equiv
{\Gamma(D \to f)-\Gamma(\bar D \to f) \over 
\Gamma(D \to f)+\Gamma(\bar D \to f)}.
\eeq
It receives both direct and indirect CP violation contributions (see, for
example, \cite{Grossman:2006jg}). 
In the SM the indirect CP violation lies well below the present
experimental sensitivity.
We therefore assume
that within the SM the measurement of  ${\cal A}_{CP}(D\to f)$ 
equals the direct CP asymmetry
\beq\label{Adir.eq}
{\cal A}_f^{\rm dir} \equiv 
{|A_f|^2 -|\bar A_{f}|^2 \over |A_f|^2  + | \bar A_{ f } |^2 }  = 2 r_f \sin \gamma \sin \delta_f. 
\eeq
Here we have used the fact that in the SM the CP-conjugate decay amplitudes for
CP even final states can be written as  
\beq
\begin{split}
A_f & \equiv  A(D \to f ) = A^T_{f} \big[1+r_f e^{i(\delta_f-\gamma)}\big],\\
\overline{A}_{f} & \equiv A(\bar D \to f ) =  A^T_{f}\big[1+r_f e^{i(\delta_f+\gamma)}\big].
\end{split}
\eeq
$A^T_{f} $ is the dominant amplitude that is proportional to
$(V_{cs} V_{us}^*- V_{cd} V_{ud}^*)$, see \eqref{eq:Heff}, and
$r_f$ is the relative magnitude of the subleading amplitude, which is proportional
to $V_{cb} V_{ub}^*$.  It carries the weak CKM phase
$\gamma=(67.3^{+4.2}_{-3.5})^\circ$~\cite{Charles:2011va} and the relative strong
phase $\delta_f $.

We perform a fit to the branching ratios and CP asymmetries to determine $r_f$ for $f=K^+ K^- \,,\,\pi^+ \pi^-$. In the fit we use the HFAG averages for the individual 
time-integrated CP
asymmetries\footnote{The average makes sense in the limit of negligible indirect CP asymmetry.} (which includes the Babar \cite{Aubert:2007if},  Belle \cite{Staric:2008rx}, 
and CDF measurements \cite{Aaltonen:2011se}),
\beq
\begin{split}\label{eq:ACPKpi} 
{\cal A}_{CP}(D\to \pi^+\pi^-) &= (0.22 \pm 0.24 \pm
0.11)\%\,, \\ 
{\cal A}_{CP}(D\to K^+K^-) &= (-0.24 \pm 0.22 \pm
0.10)\%\,, 
\end{split}
\eeq
and their difference measured at LHCb~\cite{Aaij:2011in}
\beq
\Delta {\cal A}_{CP}={\cal A}_{CP}(D\to K^+K^-) -
{\cal A}_{CP}(D\to \pi^+\pi^-) = (-0.82 \pm 0.21 \pm
0.11)\%\, ,
\eeq
and CDF ~\cite{CDF-talk}
\beq \label{eq:CDFACP}
\Delta {\cal A}_{CP}= (-0.62 \pm 0.21 \pm
0.10)\%\,.
\eeq


For strong phases $\delta_f \sim {\mathcal O}(1)$ we have 
 \beq
 \Delta {\mathcal A}_{CP}\sim 4 r_f,
 \eeq
 using the fact that $\sin\gamma\sim 0.9$, and the U-spin based expectation that ${\cal A}_{CP}(D\to K^+K^-)$ and
${\cal A}_{CP}(D\to \pi^+\pi^-)$ have opposite signs.  In order to explain the central values of $\Delta {\mathcal A}_{CP}$ one needs 
\beq\label{rf}
r_f\sim 0.2 \% \, ,
\eeq
or, equivalently, 
\beq\label{PovT}
{P /T} \sim 3  \,,
\eeq
after accounting for CKM factors.  In \cite{Brod:2011re} the penguin
contraction contributions were estimated to yield $r_f \sim 0.1\% $, 
(or $P /T \sim 1.6$), with a factor of a few uncertainty. This
motivates us to regard a hierarchy for $P/T$ that is much
larger than \eqref{PovT} as unlikely.

Our main point is that under the assumption of nominal U-spin
breaking, a broken penguin $P_{\rm break}$ which explains the
difference of the $D^0 \to K^+K^-$ and $D^0 \to \pi^+\pi^-$ decay
rates implies a $\Delta U=0$ penguin $P$ that naturally yields
\eqref{rf} and the observed $\Delta {\mathcal A}_{CP}$. 
The scaling $P_{\rm break } \sim \epsilon_U P$ together with our fit
result $P_{\rm break } \sim T/2$ (see Fig.~\ref{FigPbT}) yields the
estimate
\beq \label{randPovT} 
r_{\pi^+\pi^-, K^+K^-}\simeq
\left|\frac{V_{cb}V_{ub}}{V_{cs}V_{us}}\right|\cdot
\left|\frac{P}{T\pm P_{\rm break}}\right|\sim
\frac{|V_{cb}V_{ub}|}{|V_{cs}V_{us}|} \frac{1}{2\,\epsilon_U}\sim 0.2\%,
\eeq
for $\epsilon_U \sim 0.2$, consistent with \eqref{rf}.

In order to exhibit this result in detail, we fit the expressions in
Eqs.~\eqref{diagrammatic-begin}--\eqref{diagrammatic-end} to the four
branching ratios and the time-integrated CP asymmetries in
\eqref{eq:ACPKpi}--\eqref{eq:CDFACP}.  The latter are identified with
the corresponding direct CP asymmetries, under our assumption of
negligible indirect CP violation in the SM.
This gives us eight measurements that are fitted using five unconstrained real
parameters: the magnitudes of $T,P, P_{\rm break}$ and the two
relative strong phases. In addition there are four U-spin breaking
parameters, $\epsilon_{T1}, \epsilon_{T2}, \epsilon_{sd}^{(2)}, \epsilon_P$ that  enter at the first subleading order. They are allowed
to lie in the range $[0,0.4]$ with arbitrary strong phases. The $\chi^2$ global
minimum is $\chi^2_\text{min}=1.14$. It is not zero
because the measurements of $\Delta {\mathcal A}_{CP}$, ${\mathcal A}_{CP}(K^+K^-)$, and
${\mathcal A}_{CP}(\pi^+\pi^-)$ are only consistent at the $\sim 1 \sigma$ level.

\begin{figure}
\includegraphics[scale=0.55]{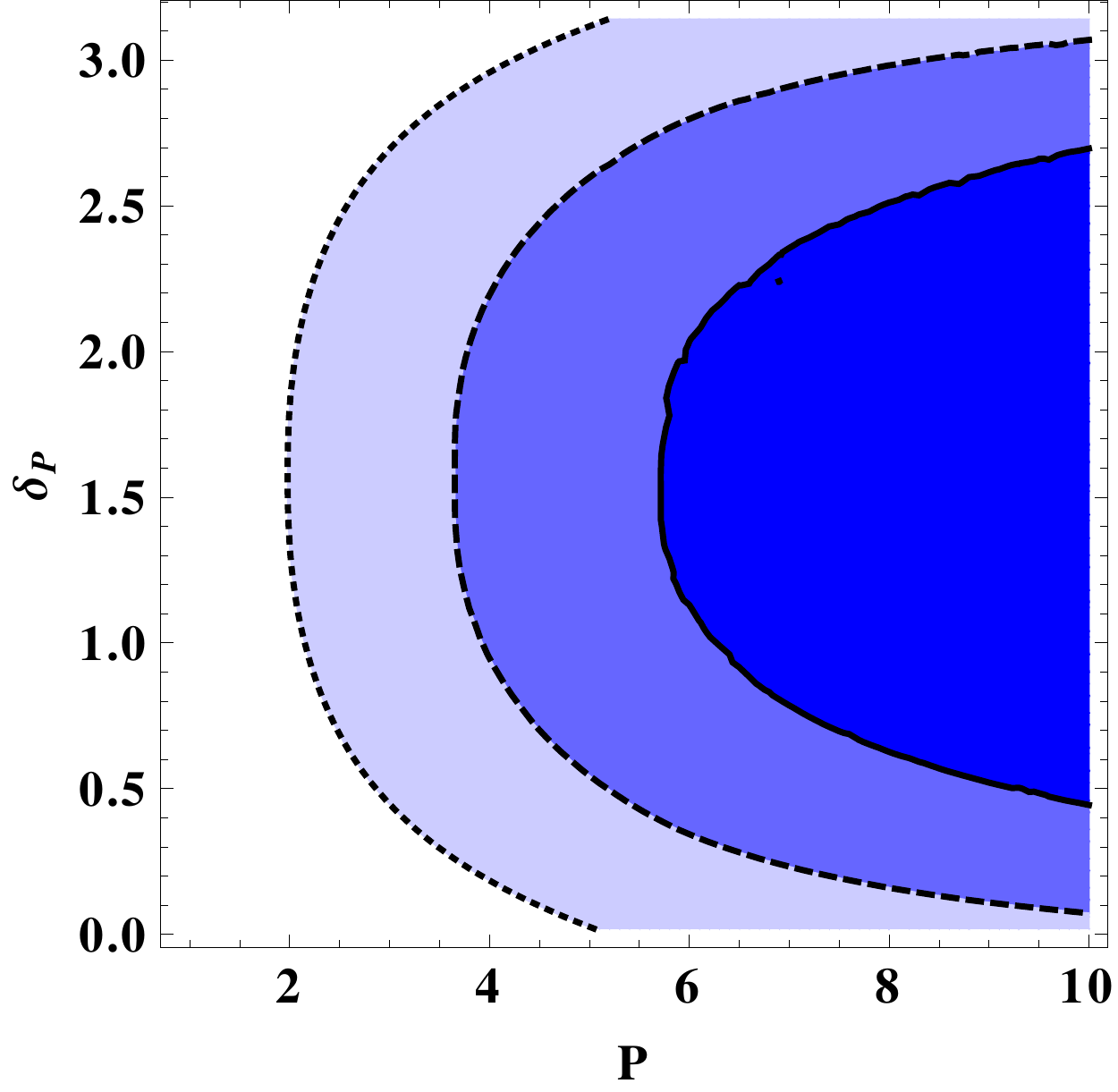}~~~~~~~~~~
\includegraphics[scale=0.55]{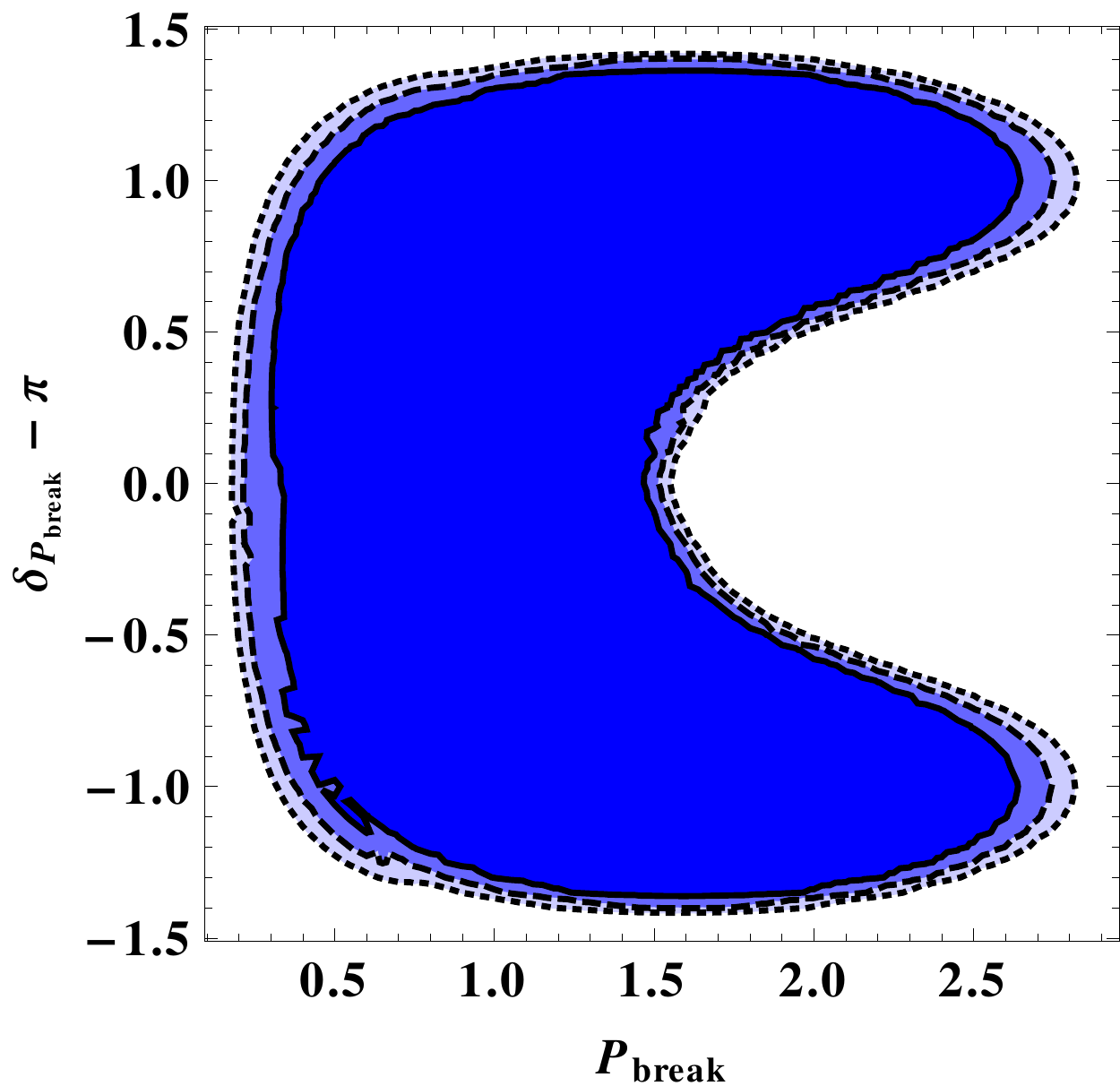}
	\caption{Constraints on $\delta_P$ vs. $P$ (left) and 
	$\delta_{P_\text{break}}$ vs. $P_\text{break}$ (right) 
	following from the fit to branching ratios and CP asymmetries 
           at $1\sigma$ (solid), $2\sigma$ (dashed), $3\sigma$
          (dotted).}
        \label{fig:PdP}
\end{figure}

In Fig.~\ref{fig:PdP} (left) 
we show the constraints on $P$ vs. the
strong phase $\delta_P\equiv\arg(P/T)$ obtained from the fit.  As
expected, small values of the penguin amplitude $P$ require a strong
phase close to $\pi/2$, whereas larger values of $P$ allow for smaller
phases. It is important to note that the minimum value of $P$ required
at $1\,\sigma$ is $\approx  5.8$, or roughly a factor of 2 larger than $T_{\rm avg}=2.83$, the
average value of $T$ in our normalization, which can be read off of
Fig.~\ref{FigPbT}. 
We also note that significant strong phases are typical for
$P_{\rm break}$, as shown in Fig.~\ref{fig:PdP}
(right), where constraints on $P_{\rm break}$ vs. the strong phase
difference $\delta_{P_{\rm break}}\equiv\arg(P_{\rm break}/T)$ are
shown.

\begin{figure}
\includegraphics[scale=0.33]{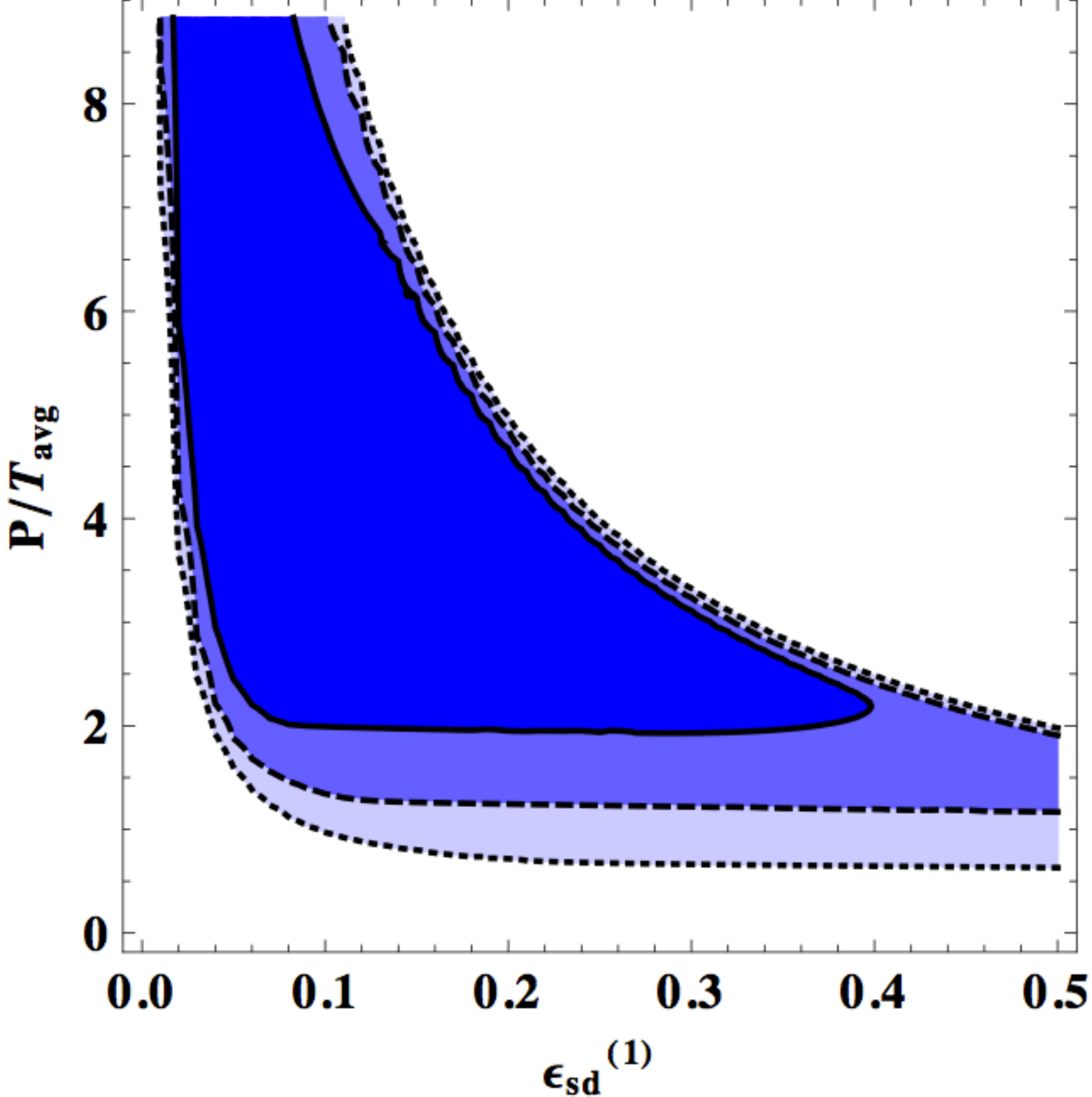}
	\caption{Constraints on $P/T_{\text{avg}}$
          vs. $\epsilon_{sd}^{(1)}$, where $T_{\text{avg}}=2.83$. The
          contours denote regions allowed at $1\sigma$ (solid),
          $2\sigma$ (dashed), $3\sigma$ (dotted). }
        \label{fig:PTepssd}
\end{figure}

\begin{figure}
\includegraphics[scale=0.33]{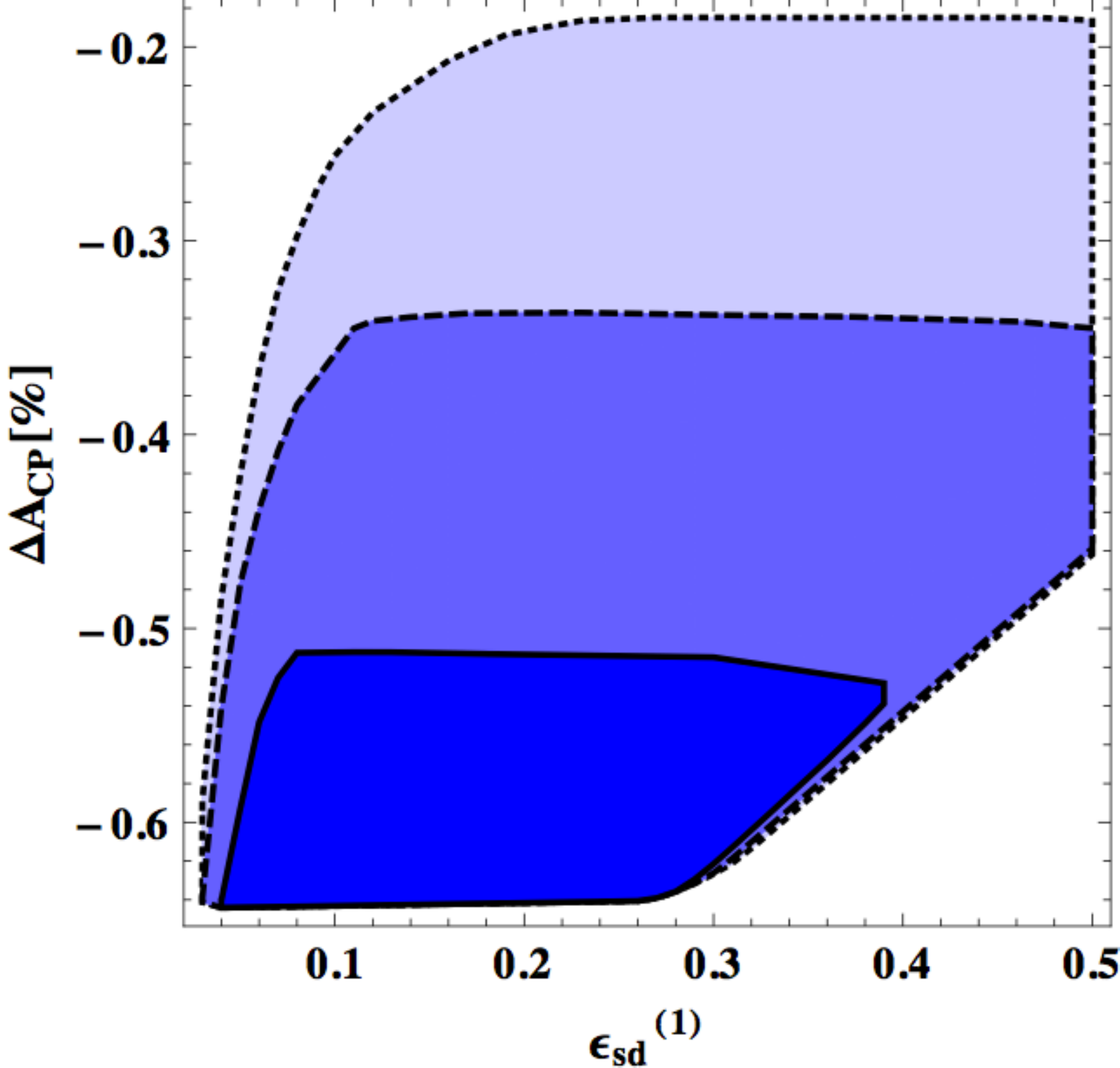}~~~~~~~~~
\includegraphics[scale=0.33]{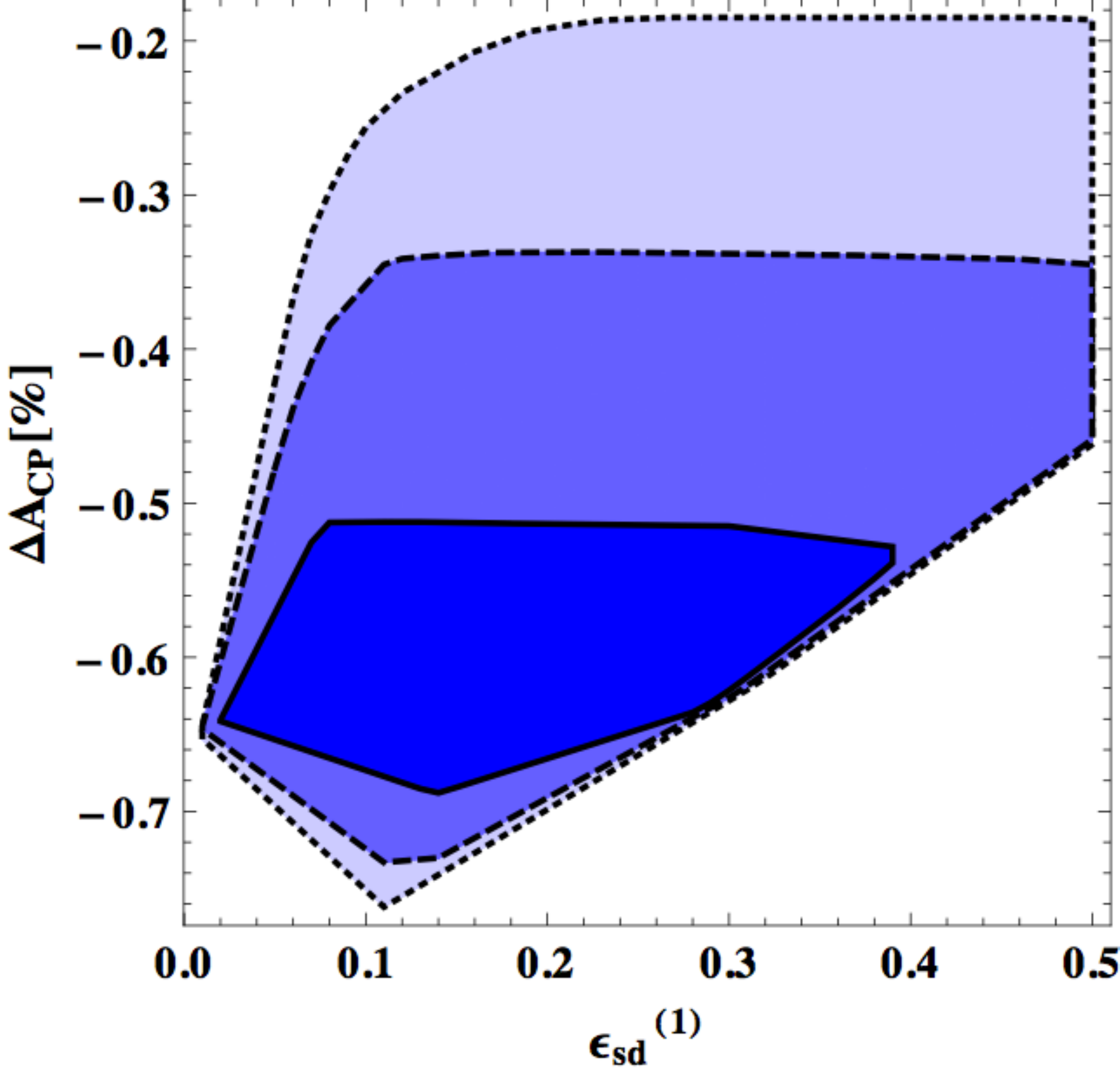}
	\caption{Allowed values of $\Delta {\mathcal A}_{CP}$
          vs. $\epsilon_{sd}^{(1)}$ corresponding to the regions shown in
          Fig.~\ref{fig:PTepssd} for
          $P\leq 10$ (left) and $P\leq 25$ (right). The contours denote regions allowed at $1\sigma$
          (solid), $2\sigma$ (dashed), $3\sigma$ (dotted). }
        \label{fig:Deltavseps10}
\end{figure}

Our main results are contained in Fig.~\ref{fig:PTepssd} and
Fig.~\ref{fig:Deltavseps10}.   We introduce the parameter $\epsilon_{sd}^{(1)}$, such that
\beq P_{\rm break } = \epsilon_{sd}^{(1)} \, P \,, \label{epsilonsd}
\eeq
as in \eqref{firstorderPb}.  
If our fit favored $\epsilon_{sd}^{(1)} $ in the nominal range for U-spin breaking,
it would support large penguins and a SM explanation for $\Delta {\cal A}_{CP}$, as in \eqref{randPovT}.

In Fig.~\ref{fig:PTepssd} the results for $P/T_{\text{avg}}$
vs. $\epsilon_{sd}^{(1)}$ are shown for an extended range, $P \leq 25$,
although it is understood that the range $P \lsim 10$ is physically
preferred. Indeed, at $1\,\sigma$ we find that $\epsilon_{sd}^{(1)}$
naturally falls into the nominal range $[0.1 ,0.3]$, for the
physically reasonable range $P/T_{\rm avg} \lsim 3.5$, equivalent to
$P\lsim 10$.  Note that the lower edges of $P/T_{\rm avg}$
correspond to the minimal values of $P$ seen in Fig.~\ref{fig:PdP}.
Higher values of $P$ are compensated by smaller values of
$\epsilon_{sd}^{(1)}$ to yield the observed difference between the
$K^+ K^-$ and $\pi^+ \pi^-$ rates, and by smaller strong phases to
yield the observed CP asymmetries.

Fig.~\ref{fig:Deltavseps10} directly addresses the question of whether we can
accommodate the CP asymmetries with nominal U-spin breaking. In
Fig.~\ref{fig:Deltavseps10} we show the values for $\Delta {\mathcal
  A}_{CP}$ for the allowed regions in Fig.~\ref{fig:PTepssd}, for the
physically more motivated range $P \lsim 10$, 
 and for an extended range of $P$. We see that we can
naturally explain the world average for $\Delta {\mathcal
  A}_{CP}=-0.67\pm 0.16$ in \eqref{DeltaACP} with nominal U-spin breaking. Note in
particular that, while values of $P>10$ allow for
marginally larger absolute values for $\Delta A_{CP}$, the lower bound
on $P$ arising from the need to explain the difference of the
branching ratios translates into a lower bound on the magnitude of
$\Delta {\mathcal A}_{CP}$ (the upper edges in
Fig.~\ref{fig:Deltavseps10}). The somewhat unexpectedly large measured
value is thus naturally explained.

\begin{figure}
	\includegraphics[scale=0.3]{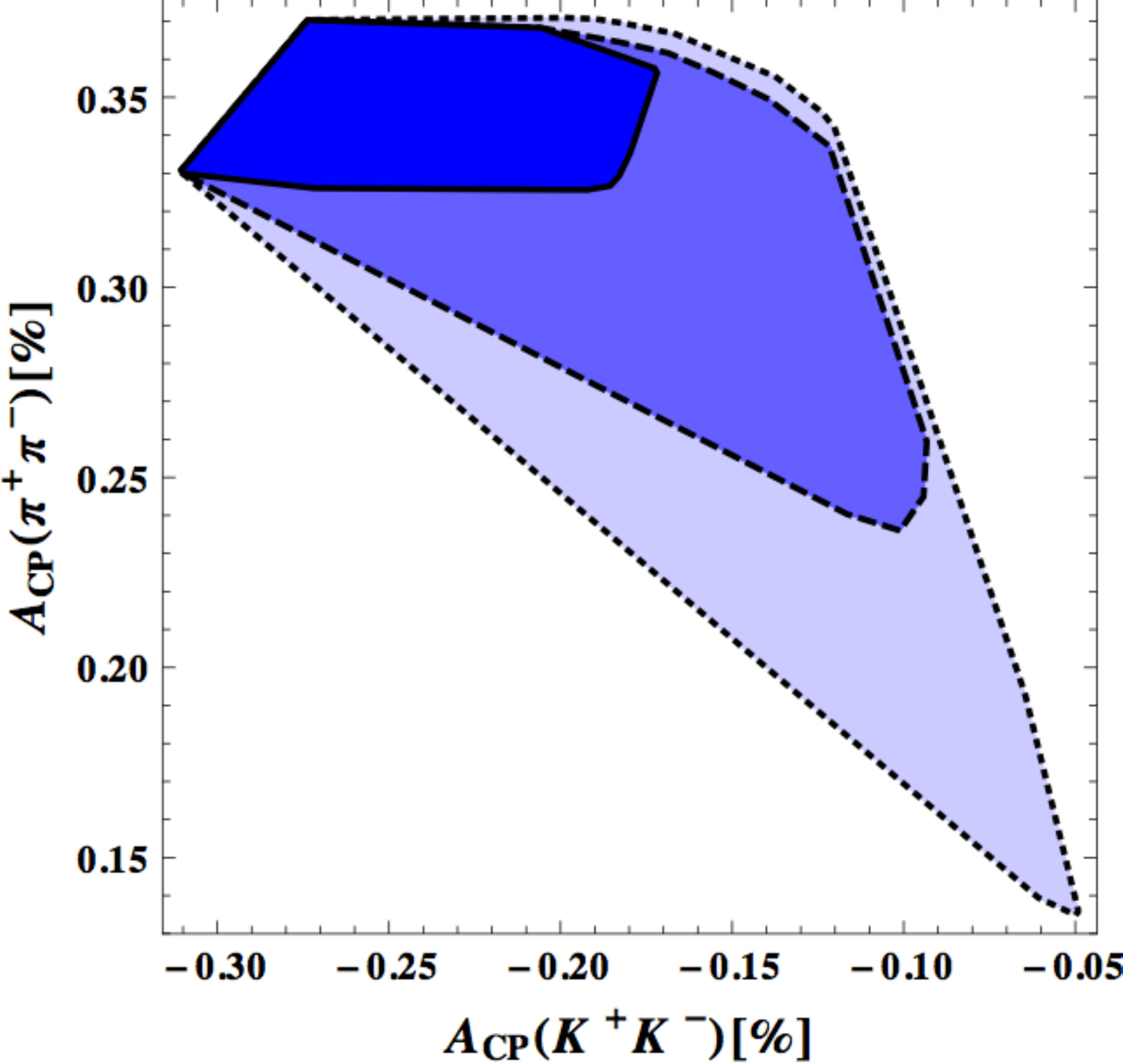}~~~~~~~~~
	\includegraphics[scale=0.3]{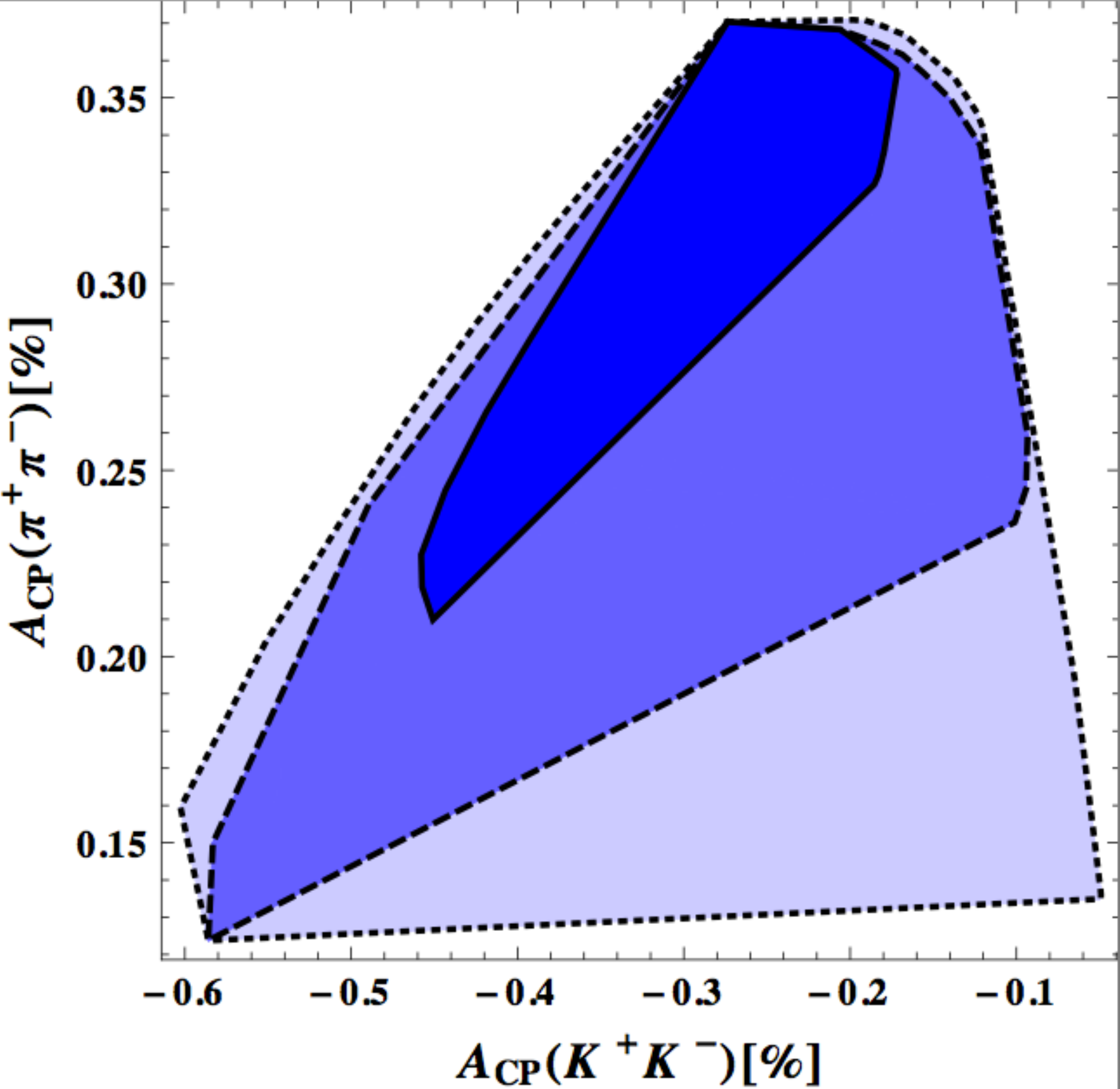}
	\caption{Result of the fit for ${\mathcal A}_{CP}(\pi^+\pi^-)$
          vs. ${\mathcal A}_{CP}(K^+K^-)$, with $P\leq 10$  (left) and
          $P\leq 25$  (right). The
          contours denote regions allowed at $1\sigma$ (solid),
          $2\sigma$ (dashed), $3\sigma$ (dotted). }
        \label{fig:acpacpPeps25}
\end{figure}

\begin{figure}
	\includegraphics[scale=0.3]{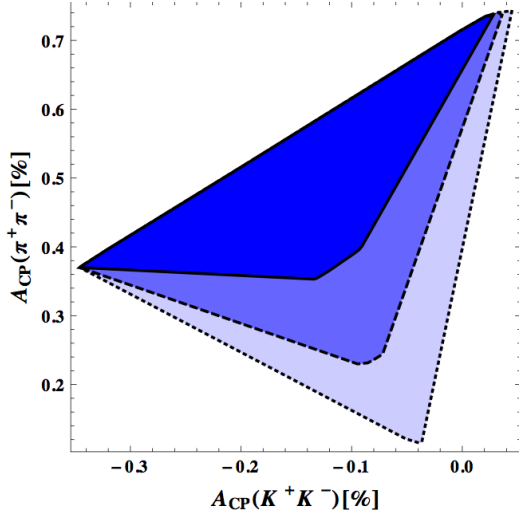}
	\caption{Result of the fit for ${\mathcal A}_{CP}(\pi^+\pi^-)$
          vs. ${\mathcal A}_{CP}(K^+K^-)$ without the individual $CPV$
          measurements, for $P\leq 10$. The contours denote regions allowed at $1\sigma$
          (solid), $2\sigma$ (dashed), $3\sigma$ (dotted). } 
        \label{fig:acpacpnoindiv}
\end{figure}

In Fig.~\ref{fig:acpacpPeps25} we show the constraints on the individual 
CP asymmetries, ${\mathcal A}_{CP}(\pi^+\pi^-)$ and ${\mathcal
  A}_{CP}(K^+K^-)$, that follow from our fit. In the U-spin
limit we would have ${\mathcal A}_{CP}(\pi^+\pi^-)=-{\mathcal
  A}_{CP}(K^+K^-)$. For nominal U-spin breaking and to ${\mathcal O}(\epsilon^\prime /\epsilon_U)$,
 we have $P_{\pi^+ \pi^- } = P_{K^+ K^- }$.  Thus, we expect the 
 asymmetries to scale like 
 \beq \label{ACPvsACP}
 {{\mathcal A}_{CP}(\pi^+\pi^-)\over {\mathcal A}_{CP}(K^+ K^-)} \approx - \left| {A(\bar D^0\to \pi^+\pi^-) \over A(\bar D^0\to K^+K^-)}\right|
 \approx -1.8 \big(1 + {\mathcal O}(\epsilon_U)\big)\,,\eeq
This is seen to
be true for the physically more motivated range $P\lsim 10$,
while for larger values of $P$ much larger values of $|{\mathcal
  A}_{CP}(K^+K^-)|$ are still allowed by the data.
Although the fit contains the individual CP asymmetry measurements as inputs, this
is a non-trivial result given that their $1\sigma $ intervals are
substantially larger than those returned by the fit. For
completeness, in Fig.~\ref{fig:acpacpnoindiv} we show the result
obtained without inputting the individual CP asymmetry measurements. A larger
hierarchy for the individual CP asymmetries becomes possible at
$1\sigma$.

Finally, let us comment on the sum rule in \eqref{sumrule-exp}, which
is fulfilled experimentally to ${\mathcal O}(4\%)$. The corresponding
amplitude-level sum rule (absolute values removed) is satisfied to $\mathcal{O}(\epsilon_U^2)$.
Taking the square roots of the branching ratios at ${\mathcal O}(\epsilon_U  /\epsilon')$, \eqref{sumrule-exp} becomes
\begin{equation}\label{eq:linsumrule}
\Sigma_{\text{sum-rule}} = \bigg(1 - \frac{1}{2}\bigg|1-\frac{P_\text{break}}{T}\bigg| -\frac{1}{2} \bigg|1+\frac{P_\text{break}}{T}\bigg|+{\mathcal O}(\epsilon_U, \epsilon_U^2/\epsilon') \bigg) = {\mathcal O}(4\%)\,. 
\eeq
Note that $\Sigma_{\text{sum-rule}} $ is not necessarily small for 
$P_{\rm break}\sim T$ and large
relative strong phase.  For instance, $|P_{\rm break}|= |T|$ yields a maximum 
for $\Sigma_{\text{sum-rule}}$ of 20\%, realized at $\delta_{P_{\rm break}}=90^\circ$.  However, the sum-rule decreases rapidly for smaller $P_{\rm break}$
and $\delta_P$.
For example, for the choices $P_{\rm break}/T  = 0.5$ (or $P_{\rm break }\approx 1.4$) and $\delta_{P_{\rm break}}=\pm 45^\circ$, which lie near the two ``foci" of the $1\sigma $ region in Fig.~\ref{fig:PdP} (right), the sum rule is reduced to $\approx 6.6\%$ at ${\mathcal O}(\epsilon_U/\epsilon')$, marginally larger than 
experiment.    We have checked that the degree of tuning  in the description of all observables considered is modest,  
about 1 part in 3, with the dominant effect due to the sum rule.  This is what one would expect 
from the  ${\mathcal O}( \epsilon_U , \epsilon_U^2 /\epsilon')$ corrections to  \eqref{eq:linsumrule},
assuming  U-spin breaking of nominal size.
These corrections would necessitate substantial tuning of
$\Sigma_{\text{sum-rule}}$ if $\epsilon_U$ were large, thus providing
further support for nominal U-spin breaking.

\section{Conclusion}\label{sec:concl}

We have shown that the penguin contraction matrix elements of the
standard-model operators $Q_{1,2}$ can provide a consistent picture
for large penguins in singly Cabibbo-suppressed $D \to PP$ decays.
The U-spin violating contractions of $Q_{1,2}$ explain the
long-standing puzzle of a significantly larger branching ratio for
$D\to K^+K^-$ compared to $D\to \pi^+\pi^-$.  At the same time, the
U-spin conserving contractions of $Q_{1,2}$ are of the correct size to
naturally explain the large CP asymmetry difference $\Delta {\cal
  A}_{CP}={\cal A}_{CP}(D\to K^+K^-) - {\cal A}_{CP}(D\to \pi^+\pi^-)
$ measured by LHCb and CDF.  A crucial observation, borne out by a
detailed U-spin analysis, is that U-spin breaking of nominal size,
${\mathcal O}(20\%)$, correctly relates the magnitudes of the two
phenomena. On this basis, we conclude that large direct CP asymmetries
of order a few per mille are not surprising given the size of
$\text{Br}(D\to K^+K^-)/\text{Br}(D\to \pi^+\pi^-)$.

\section*{Acknowledgements}
We thank Gino Isidori and Gilad Perez for useful discussions.
Y.~G. is supported in part by the NSF grant PHY-0757868 and by a grant
from the BSF. J.~B. and A.~K. are supported by DOE grant
FG02-84-ER40153.  This work was facilitated in part by the workshop
"New Physics from Heavy Quarks in Hadron Colliders" which was
sponsored by the University of Washington and supported by the DOE
under contract DE-FG02-96ER40956.

\appendix
 \section{Formal U-spin decomposition}
 \label{Formal-U-spin-app}

In this appendix we perform the U-spin decomposition of the $D\to
P^+P^-$ decays ($P=K,\pi$), including U-spin breaking.  The most
general expressions for the amplitudes are truncated at second order
in U-spin breaking, that is, there are no new hadronic matrix elements
introduced at higher orders.  The U-spin decomposition is written in a
form in which it is clear which reduced matrix elements contain the
penguin contractions.  We assume that these matrix elements are
dynamically enhanced.  A translation is provided between the
amplitudes in the U-spin decomposition and the decomposition given in
Section II.

\subsection{General discussion}
The $D^0$ meson is a U-spin singlet. The decay operators involve a
down-type quark and a down-type antiquark, and thus in general they can be
written as a sum of a U-spin triplet and a U-spin singlet. The leading
operators~\eqref{eq-t1} form a triplet, while the
U-spin singlet operator is proportional to $V_{cb}V_{ub}^*$. 
It is completely negligible as far as the CP
averaged rates are concerned, due to the CKM suppression.  In terms of tensor notation the two
operators are given by 
\beq \label{Uspinops} H_1= \langle H_1 \rangle
\begin{pmatrix} \tfrac{1}{2}\big(V_{cs}V_{us}^*- V_{cd}V_{ud}^*\big) &
V_{cs}^*V_{ud} \\ V_{cd}V_{us}^* & -\tfrac{1}{2}\big( V_{cs}V_{us}^*-
V_{cd}V_{ud}^* \big) \end{pmatrix},\qquad  
 H_0=  - V_{cb}^*V_{ub}\, \langle H_0 \rangle \,{\rm I}_{2 \times 2},
\eeq
where the $\langle H_i\rangle$ are hadronic coefficients, and $H_0$ is
proportional to the identity matrix ${\rm I}_{2 \times 2}$. 
We also split the final states into a U-spin triplet and singlet 
\beq \label{Mfinalstates}
M_1= \begin{pmatrix}
\tfrac{1}{2}\big( K^+K^--\pi^+\pi^-\big) & \pi^+ K^- \\
 \pi^- K^+ & -\tfrac{1}{2}\big(K^+K^--  \pi^+\pi^-\big)
 \end{pmatrix},
 \qquad
M_0=
\tfrac{1}{2}\big( \pi^+\pi^-+K^+K^-\big)\, {\rm I}_{2 \times 2}.
\eeq 
The U-spin breaking is induced by the nonzero strange-quark mass
term, $m_s \bar s s$.  Subtracting the singlet piece, the breaking
introduces a spurion that is a U-spin triplet
\beq
M_\epsilon=
 \begin{pmatrix}
\epsilon/2 & 0\\
0& -\epsilon/2
\end{pmatrix}\,,
 \eeq
where $\epsilon$ is a small parameter which
parametrizes U-spin breaking, i.e., $\epsilon \sim \epsilon_U$.

In the U-spin limit there are two reduced matrix elements
\beq \label{eps0}
t_0 \propto \langle f_1|H_1|0\rangle, \qquad
p_0 \propto \langle f_0|H_0|0\rangle,
\eeq
where $f_0 $ and $f_1 $ are the singlet and triplet states
corresponding to $M_0 $ and $M_1$, respectively, and $|0\rangle$ is
the U-spin singlet $D^0$ meson.
At ${\mathcal O}(\epsilon)$ there are three additional reduced matrix
elements
\beq \label{eps1}
s_1\propto \langle f_0|(H_1\times \1)_0|0\rangle, \qquad
t_1\propto \langle f_1|(H_1\times \1)_1|0\rangle, \qquad
p_1\propto \langle (f_1 \times \1 )_0 | H_0|0\rangle, \qquad
\eeq
where $\1$ represents the U-spin breaking spurion
$M_\epsilon$.
At ${\mathcal O}(\epsilon^2)$ there are three more reduced matrix elements,
\beq\begin{split}\label{eps2}
t_2 \propto   \langle (f_1 \times \1)_0 & |(H_1\times\1)_0   |0\rangle \,,\qquad
t_2' \propto  \langle (f_1 \times \1)_1 |(H_1\times\1)_1   |0\rangle\,,\\
&~~~p_2 \propto  \langle f_0|H_0\times(\1\times \1)_0 |0\rangle.
\end{split}
\eeq

In terms of the tensor notation, we have the identities
\beq
\begin{split}   \label{H1singlet}
(H_1\times \1)_0 \equiv &\, \{H_1, M_\epsilon \} =  \epsilon\, \langle H_1 \rangle \tfrac{1}{2}\big(V_{cs}V_{us}^*- V_{cd}V_{ud}^*\big)\,  {\rm I}_{2 \times 2}
 \\
(H_1\times \1)_1 \equiv &\, [H_1, M_\epsilon ] =  \epsilon \, \langle H_1 \rangle \begin{pmatrix} 0 &
-V_{cs}^*V_{ud} \\  V_{cd}V_{us}^* & 0 \end{pmatrix}.
\end{split}
\eeq
Thus, the anticommutator (commutator) of $H_1 $ and $M_\epsilon$ projects onto a singlet (triplet) operator.
Similarly, we have
\beq
\begin{split}   \label{f1singlet}
(f_1\times \1)_0  \equiv &\, \{M_\epsilon, M_1 \} = \epsilon\, \tfrac{1}{2} \big( K^+K^--\pi^+\pi^-\big) \,  {\rm I}_{2 \times 2}
 \\
(f_1\times \1)_1  \equiv &\, [M_\epsilon, M_1 ] = \epsilon\,  \begin{pmatrix} 0 &
\pi^+ K^-  \\ - \pi^- K^+ & 0 \end{pmatrix},
\end{split}
\eeq
and the anticommutator (commutator) of $M_1 $ and $M_\epsilon$ projects onto a singlet (triplet) state.

We can now write the two-body decay Hamiltonian to ${\mathcal
O}(\epsilon^2)$ as 
\beq
\begin{split}\label{H1Trace}
{\cal H}_1=&t_0 \tr(H_1M_1)+\tfrac{1}{2}t_1 \tr([H_1,M_\epsilon] M_1)+\tfrac{1}{4}t_2 \tr(\{H_1, M_\epsilon\} \{M_\epsilon, M_1\})\\
+&\tfrac{1}{4}t_2' \tr([H_1, M_\epsilon] [M_\epsilon ,M_1 ])+s_1\tr(\{H_1, M_\epsilon\} M_0  ),
\end{split}
\eeq 
for the U-spin triplet operator, and 
\beq
\begin{split}\label{H0Trace}
{\cal H}_0=&p_0 \tr(H_0 M_0)+\frac12 p_1 \tr(H_0 \{M_\epsilon ,M_1 \})+p_2 \tr(H_0 M_\epsilon^2 M_0).
\end{split}
\eeq 
for the U-spin singlet operator.

Choosing a convenient final state
phase convention, the decay amplitudes can be read off from \eqref{H1Trace}, \eqref{H0Trace}, yielding 
\beq\label{U-spin-decomp}
\begin{split}
A(\bar D^0\to K^+\pi^-)&=V_{cs} V_{ud}^* \left( t_0 -\tfrac{1}{2} t_1 \epsilon
+ \tfrac{1}{4} t_2' \epsilon^2\right) ,\\ 
A(\bar D^0\to \pi^+\pi^-)&=
-\tfrac{1}{2}\big(V_{cs}V_{us}^*-
V_{cd}V_{ud}^*\big)\left(t_0+s_1\epsilon +\tfrac{1}{2} t_2
\epsilon^2\right) -V_{cb}V_{ub}^*\left(p_0-\tfrac{1}{2}p_1\epsilon
+\tfrac{1}{4}p_2\epsilon^2\right),\\ 
A(\bar D^0\to K^+K^-)&=
\tfrac{1}{2}\big(V_{cs}V_{us}^*- V_{cd}V_{ud}^*\big)\left(t_0-
s_1\epsilon +\tfrac{1}{2} t_2\epsilon^2\right) - V_{cb}V_{ub}^*
\left(p_0+\tfrac{1}{2}p_1\epsilon
+\tfrac{1}{4}p_2\epsilon^2\right),\\ 
A(\bar D^0\to \pi^+K^-)&=V_{cd} V_{us}^*
\left(t_0+\tfrac{1}{2}t_1\epsilon+\tfrac{1}{4}t_2' \epsilon^2
\right) ,
\end{split}
\eeq
where we have made the replacements $t_i \langle
H_1\rangle \to t_i$ and $p_i \langle H_0\rangle \to p_i$ for the reduced
matrix elements.

The U-spin expansion is just a basis rotation, so that in general there can only be six independent decay 
amplitudes, four associated with decays to the triplet final state $f_1$ and two with decays to the singlet final state $f_0$.
Thus, of the three reduced matrix elements introduced at ${\mathcal O}(\epsilon^2)$, only one combination is a new linearly independent amplitude.  In fact, we can see directly from the above amplitude expressions that $p_2$ can be absorbed into $p_0$, and that one linear combination of $t_2 $ and $t_2 '$ can be absorbed into $t_0$, leaving another linear combination of the two as the new linearly independent amplitude.

There are a number of conclusions that one can draw from the above
decomposition at different orders in $\epsilon$. In the U-spin
symmetric limit we have the following known relations:
(i) all 4 decay rates are equal (up to CKM prefactors); and
(ii) the direct CP asymmetries in the SCS decays are equal in
magnitude and opposite in sign. 

Working to ${\mathcal
O}(\epsilon)$, and neglecting the small terms proportional to
$V_{cb}V_{ub}^*$, the four amplitudes depend on three reduced matrix elements.
Thus, there is one relation among the $\bar D^0$ decay amplitudes, which is given by
\beq \label{sumrule-the}
\frac{\bar{A}_{K^-\pi^+}}{V_{cs} V^*_{ud}}+\frac{\bar{A}_{K^+\pi^-}}{V_{cd} V^*_{us}} = \frac{\bar{A}_{K^+ K^-}}{V_{cs} V^*_{us}}+\frac{\bar{A}_{\pi^+ \pi^-}}{V_{cd} V^*_{ud}},
\eeq
and similarly for the CP conjugate decays.
This sum rule is broken at ${\mathcal
O}(\epsilon^2)$, whereas the individual amplitudes are ${\mathcal
O}(1)$. The experimental relation~\eqref{sumrule-exp} is also satisfied to first order, that is,
$\Sigma_{\text{sum-rule}} = {\mathcal O}(\epsilon^2)$.  

The rate difference between the $K^+\pi^- $ and $K^- \pi^+ $ modes and
the rate difference between the $K^+ K^- $ and $\pi^+ \pi^- $ modes
arise at order ${\mathcal O}(\epsilon)$.  However, the latter is
observed to be ${\mathcal O}(1)$, while the former is small. The
immediate conclusion is that
\beq
s_1 \epsilon  \gg t_1  \epsilon.
\eeq
There are two ways in which this relation could be realized.  The
first one is that there is a U-spin breaking hierarchy which remains
unexplained, i.e., a hierarchy of $\epsilon$'s, such that the
breaking is much larger for $s_1$ than for $t_1$.  The second
possibility, which is the one we have pursued in this paper, is that
all U-spin breaking is of nominal size, but that $s_1$ is enhanced
relative to $t_1$ due to the penguin contractions.

\subsection{The penguin contractions}

Our working assumption is that at a given order in $\epsilon$
the contractions of the $s\bar s$ and $d\bar d$ fields give 
the dominant effects.
The reduced matrix elements to which the contractions contribute are the ones in which both the transition Hamiltonian and the 
final state can be written as U-spin singlets.  
Thus, they are identified with the following traces
\beq \label{contractionrule} \tr [(M_0~{\rm or}~ \{M_\epsilon , M_1 \}) \,\times\,(H_0~{\rm or}~ \{H_1 , M_\epsilon \})] \,.\eeq
According to \eqref{H1Trace} and \eqref{H0Trace},  these are the $H_0$ matrix elements $p_0 , p_1 ,p_2$ and the $H_1$ matrix elements $s_1 , t_2$.  
We elaborate below, and check the consequences for the U-spin decomposition.

At ${\mathcal O}(\epsilon^0)$, $H_0$ gives rise to 
the reduced matrix element $p_0$.  This involves matrix elements of $Q_{1,2}^{\bar s s}+Q_{1,2}^{\bar d d}\,$ for the singlet final state $( |K^+ K^-  \rangle  + |\pi^+ \pi^- \rangle )/\sqrt{2} $, see \eqref{Uspinops} and \eqref{eq:Heff}.  
Therefore, $p_0$ contains both contracted and non-contracted contributions. Also at ${\mathcal O}(\epsilon^0)$,
the $U_3 = 0 $ component of $H_1$ gives rise to the reduced matrix element $t_0$. 
It involves matrix elements of $Q_{1,2}^{\bar s s}-Q_{1,2}^{\bar d d}$ for the triplet final state $( |K^+ K^-  \rangle  - |\pi^+ \pi^- \rangle )/\sqrt{2} $.
Therefore, the $s\bar s $ and $d\bar d$ contractions must cancel in $t_0$ at ${\mathcal O}(\epsilon^0)$.

To account for the dominance of the contractions we introduce a second small parameter
$\epsilon'$, such that any matrix element to which they contribute is enhanced by ${\mathcal O}(1/\epsilon' )$. 
Without loss of generality, we can define it as 
\beq    
\epsilon' \equiv \left|{t_0\over p_0}\right|.
\eeq
Thus, in $p_0$ the ratio of non-contracted to contracted contributions
is also of ${\mathcal O}(\epsilon' )$.

The contractions enter the matrix elements of $H_1$ for the first time at
${\mathcal O}(\epsilon)$. This happens for the $s_1$ matrix element
for which the final state is a singlet. 
This is because the operator 
 $(H_1\times \1)_0  \equiv \{H_1 , M_\epsilon\} $
contains the sum  $Q_{1,2}^{\bar s s}+Q_{1,2}^{\bar d d}$, see \eqref{H1singlet} and \eqref{eq:t3}.
Thus, the contractions do not cancel, but add up.  This is also the case for the contractions in $p_1$ at ${\mathcal O}(\epsilon)$, and the contractions in $p_2 $ and $t_2 $ at ${\mathcal O}(\epsilon^2 )$.  Note, however, that $p_1$ and $t_2$ involve the $U_3 = 0$ triplet final state. We can transform it to a singlet with the aid of the U-spin breaking spurion i.e., $\{M_\epsilon , M_0\}$.  
We are led to the following 
$\epsilon'$ counting rule: a reduced matrix element that is associated
with one of the traces in \eqref{contractionrule} is ${\mathcal
  O}(1/\epsilon')$, and is ${\mathcal O}(1)$ otherwise. 
Explicitly, 
\beq
s_1\sim t_2\sim p_{0,1,2}\sim {\mathcal O}(1/\epsilon'), \qquad t_{0,1}\sim t_2'\sim {\mathcal O}(1).
\eeq

Taking $\epsilon' \sim \epsilon \sim \epsilon_U$, the different amplitudes are thus
\beq
\begin{split}
&p_0\sim {\mathcal O}(1/\epsilon), \qquad t_0\sim s_1 \epsilon\sim p_1\epsilon\sim  {\mathcal O}(1), \qquad\\
& t_1\epsilon\sim t_2\epsilon^2\sim p_2\epsilon^2\sim {\mathcal O}(\epsilon), \qquad t_2'\epsilon^2\sim {\mathcal O}(\epsilon^2).
\end{split}
\eeq
Keeping only terms to order ${\mathcal O}(\epsilon)\sim {\mathcal O}(\epsilon^2/\epsilon')\sim {\mathcal O}(\epsilon_U)$, i.e. the leading and subleading terms for both CKM structures, we finally have
\beq\label{U-spin-decomp-2}
\begin{split}
A(\bar D^0\to K^+\pi^-)&=V_{cs} V_{ud}^* \left( t_0 -\tfrac{1}{2} t_1 \epsilon \right) ,\\
A(\bar D^0\to \pi^+\pi^-)&= -\tfrac{1}{2}\big(V_{cs}V_{us}^*- V_{cd}V_{ud}^*\big)\left(t_0+s_1\epsilon +\tfrac{1}{2} t_2  \epsilon^2\right) - V_{cb}V_{ub}^*\left(p_0-\tfrac{1}{2}p_1\epsilon \right),\\
A(\bar D^0\to K^+K^-)&= \tfrac{1}{2}\big(V_{cs}V_{us}^*- V_{cd}V_{ud}^*\big)\left(t_0- s_1\epsilon +\tfrac{1}{2} t_2\epsilon^2\right) - V_{cb}V_{ub}^* \left(p_0+\tfrac{1}{2}p_1\epsilon \right),\\
A(\bar D^0\to \pi^+K^-)&=V_{cd} V_{us}^* \left(t_0+\tfrac{1}{2}t_1\epsilon \right).
\end{split}
\eeq
Note that this is the most general decomposition, that is, all
the subleading terms that we neglected can be absorbed into terms we
kept. The decomposition in
\eqref{U-spin-decomp-2} is equivalent to the decomposition given in
\eqref{diagrammatic-begin}-\eqref{diagrammatic-end}, with the
following translations: at ${\mathcal O}(1/\epsilon)$, and also
including the ${\mathcal O}(1)$ non-contraction term in $p_0$,
\beq
p_0=P + \tfrac{1}{2} T\, ;
\eeq
at ${\mathcal O}(1)$, and also including the ${\mathcal O}(\epsilon)$ non-contraction term in $s_1$,
\beq
\begin{split}
p_1\epsilon=P\epsilon_P, 
\qquad t_0=T, \qquad s_1\epsilon=P_{\rm break} +
\tfrac{1}{2}\epsilon_{T1} T\, ;
 \end{split}
 \eeq
and at ${\mathcal O}(\epsilon)$
\beq
\begin{split}
 t_1\epsilon=T\epsilon_{T2}, \qquad t_2 \epsilon^2= - P_{\rm break} \epsilon_{sd}^{(2)}.
 \end{split}
 \eeq

Fitting \eqref{U-spin-decomp-2} to the four branching ratios, dropping the $V_{cb} V_{ub}^* $ terms, and taking $\epsilon \in [0,0.4]$ yields the fit for $s_1 \epsilon$ vs. $t_0$ shown in Fig.~\ref{Figs1t0}.
It is similar to the fit for $P_{\rm break}$ vs. $T$ in Fig.~\ref{FigPbT}, as one would expect expect,
and confirms that $s_1 \epsilon  \sim t_0 $ and $\epsilon' \sim \epsilon$.

\begin{figure}
	\includegraphics[scale=0.33]{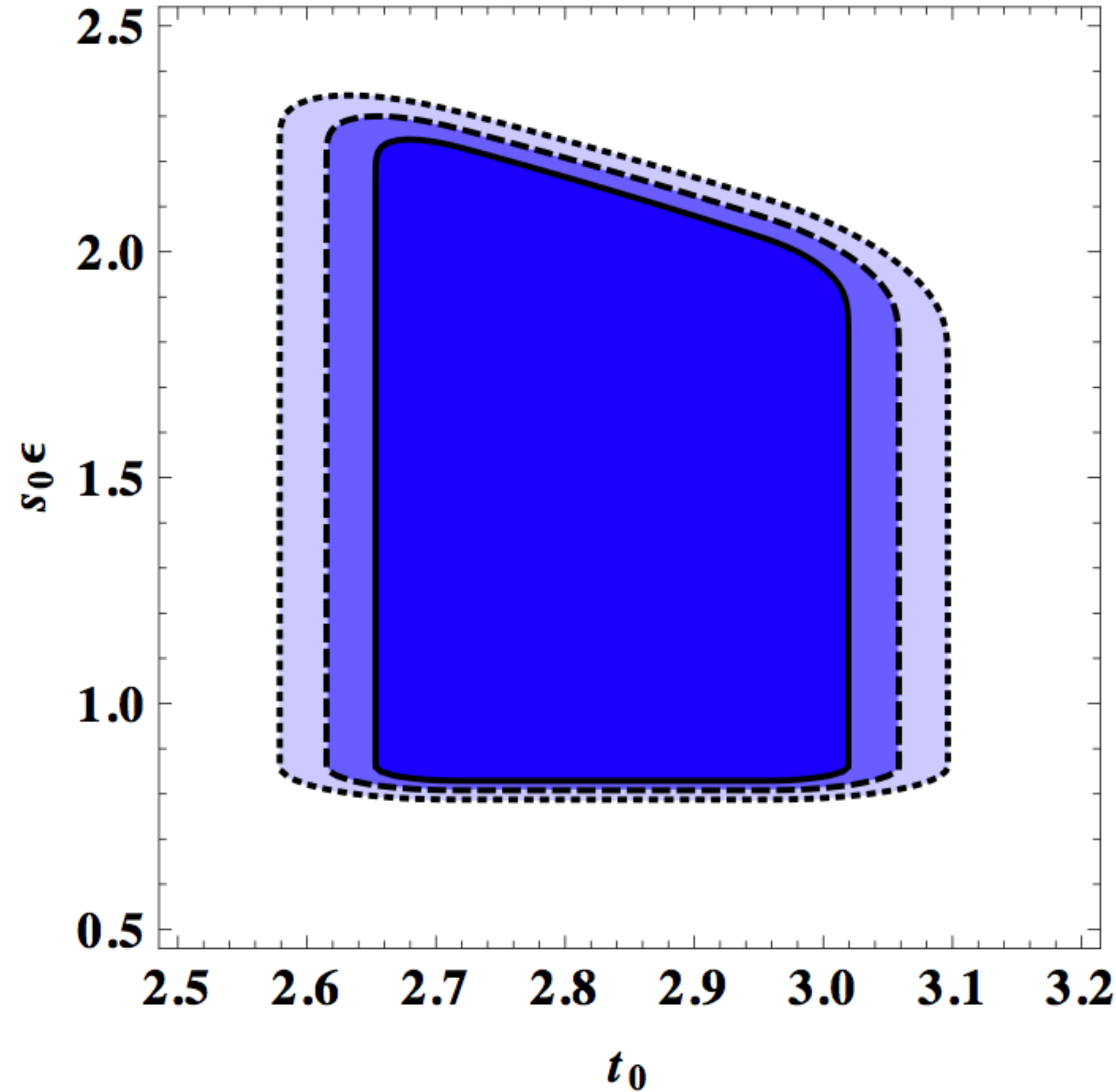}
	\caption{The results of the fit of the U-spin decomposition in \eqref{U-spin-decomp-2} to the $D\to K^+\pi^-$, $K^+K^-$, $\pi^+\pi^-$,$\pi^+K^-$ branching ratios. The contours denote the regions allowed at $1\sigma$ (solid), $2\sigma$ (dashed), $3\sigma$ (dotted).}\label{Figs1t0}
\end{figure}

\section{Diagrammatical representation}
\label{App:diagrammatics}
Finally, we provide a derivation of the U-spin decomposition presented in \eqref{diagrammatic-begin}-\eqref{diagrammatic-end} in terms of the
diagrammatical representation. This will simplify the comparison
with the result of Ref.~\cite{Brod:2011re}, and provides further insight
into the origin of the scaling $\epsilon' \sim \epsilon_U$.

The $\bar D^0$ decay amplitudes $\bar A_f$ are split into ``tree-level" amplitudes $\bar A_f^{\,T}$ and penguin amplitudes $\bar A_f^{\,P}$. The 
former are sums of a ``tree" diagram $T$ and an ``exchange" diagram $E$, and are given in full generality by
\be
\begin{split}
\bar A^{\,T}_{K^+ \pi^- }&  =  V_{cs} V_{ud}^* ( T_{K\pi} + E_{K\pi}), \\
\bar A^{\,T}_{\pi^+ \pi^-} &= -\tfrac{1}{2}\left(V_{cs} V_{us}^* -V_{cd} V_{ud}^*\right) \left( T_{\pi\pi} + E_{\pi\pi}\right),\\
\bar A^{\,T}_{K^+ K^-} &= \tfrac{1}{2}\left(V_{cs} V_{us}^* -V_{cd} V_{ud}^*\right) \left( T_{KK} + E_{KK}\right),\\
\bar A^{\,T}_{K^- \pi^+ }&  =  V_{cd} V_{us}^* ( T_{\pi K} + E_{\pi K}). 
\label{eq:Tamps}
\end{split}
\eeq 
The $T$ amplitudes are those with a $u$ spectator quark, and the $E$ amplitudes are the annihilation topology 
diagrams (cf. Fig.~\ref{fig:T}), in which the initial $c \bar u$ quark pair annihilates into $\bar s d$, $\bar s s $,  $\bar d d$, and $\bar d s$ quark pairs, respectively. 

In SCS decays we can divide $E$ and $T$ into two contributions: those that do not involve the $\bar s s$ or $\bar d d$ penguin contractions, denoted $T_f$ and $E_f$, and those which do,
denoted $P_f^T $ and $P_f^E$ and shown in the last two diagrams of Fig.~\ref{fig:T}. The penguin contractions arise from  $Q_{1,2}^{\bar s s} - Q_{1,2}^{\bar d d}$ and are U-spin violating.  At the quark level they correspond to the difference 
between the rescattering contributions of the $ \bar s s$ and $ \bar d d$ quark pairs for a given final state. We thus have
\beq
\begin{split}
T_{KK} &= T_{KK}^{s}    -\big(P_{KK}^{T,d} - P_{KK}^{T,s} \big),\qquad T_{\pi\pi} = T_{\pi\pi}^{d}  
+ \big(P_{\pi\pi}^{T,d} -P_{\pi\pi}^{T,s}\big),\\
E_{KK} &= E_{KK}^{s}  -\big(P_{KK}^{E,d} - P_{KK}^{E,s} \big),\qquad E_{\pi\pi} = E_{\pi\pi}^{d}   
+\big(P_{\pi\pi}^{E,d} -P_{\pi\pi}^{E,s} \big),
\label{eq:TandE}
\end{split}
\eeq
where the superscripts $s$ and $d$ denote the identity of the relevant operators $Q_{1,2}^{\bar s s}$ or $Q_{1,2}^{\bar d d} $ (or the identity of the contracted quark pair in the penguin contractions). In the CF and DCS decays $T$ and $E$ of course do not receive contributions from the penguin contractions.

Formally, the $P^T$ contain leading power as well as power correction contributions, while the $P^E$ are pure power corrections. 
The scheme-dependent coefficients and leading $\log\mu$ scale dependence entering the penguin contraction matrix elements cancels in the differences $P_f^{T,s} - P_f^{T,d}$ 
and $P_f^{E,s} - P_f^{E,d}$.  

\begin{figure}[t]
\centering
$\begin{array}{ccc}
\includegraphics[width=5.0cm]{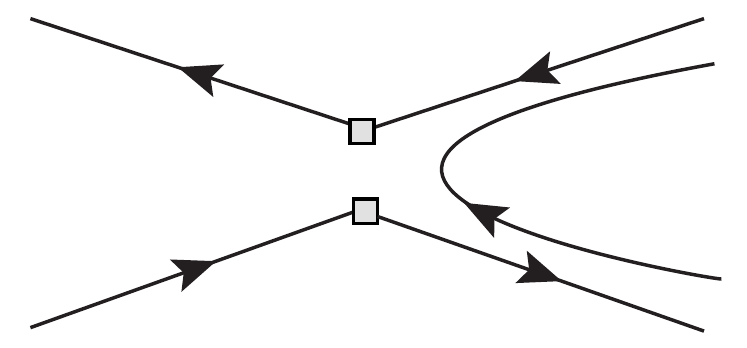} &
\includegraphics[width=5.0cm]{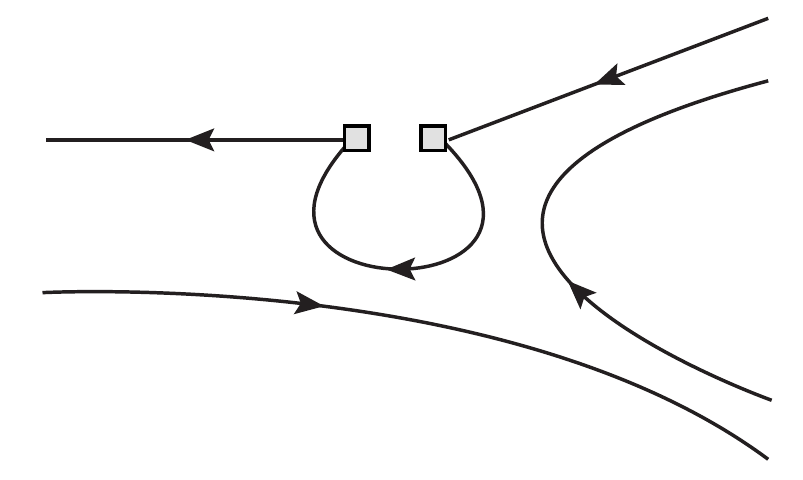} &
\includegraphics[width=5.0cm]{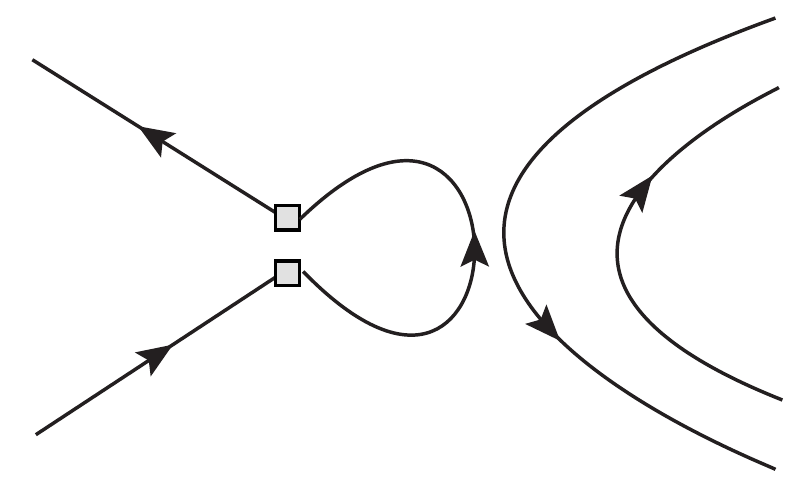}
\end{array}$
\caption{From left to right: tree level exchange topology diagram, $E_f$, and the penguin contraction diagrams $P_f^T$ and $P_f^E$. \label{fig:T} }
\end{figure}

In the penguin amplitudes the penguin contractions correspond to the sum of the rescattering contributions of the $ \bar s s$ and $ \bar d d$ quark pairs in $( Q_{1,2}^{s} + Q_{1,2}^d )/2$. 
Including the amplitudes for the penguin operators $Q_{3,..,6}$, $Q_{8g}$, denoted $\tilde P$, and the above contracted and non-contracted amplitudes, we have
\beq
\begin{split}
A^{P}_{f} &= -V_{cb} V_{ub}^*   \big(  \tilde P_{f} +  [ P_{f}^{T,s}  + P_f^{T,d} ]/2  +  [ P_{f}^{E,s}  + P_f^{E,d} ]/2  + [ T^q_f + E^q_f ]/2 \big),
\label{eq:Tamps2}
\end{split}
\eeq 
where $q=s\,(d)$ for $f = K^+ K^- \,(\pi^+ \pi^- )$.  The scheme and scale dependence in the penguin contractions is now canceled by $\tilde P_f$.  However, since the contributions of the $\tilde P_f$ to the direct CP asymmetries are much smaller than observed, they are neglected in our fits. 

To simplify the comparison with the previous section and with Section~\ref{sec:CP}, let us define the total non-contracted amplitudes,
\beq \label{noncontracted}
\begin{split}
{\mathcal T}_{K\pi} &\equiv T_{K\pi} + E_{K\pi},\qquad  {\mathcal T}_{\pi  K} \equiv T_{\pi  K} + E_{\pi K}\\
{\mathcal T}_{\pi\pi} &\equiv T^d_{\pi\pi} + E^d_{\pi\pi},\qquad {\mathcal T}_{KK} \equiv T^s_{KK} + E^s_{KK}\,,
\end{split}
\eeq
and the total penguin contraction contributions in the tree and penguin amplitudes, 
($f= K K$ ($\pi\pi$) for the $K^+ K^-$ ($\pi^+ \pi^- $) final state)
\beq
\begin{split} \label{contracted}
{\mathcal P}^t_f & \equiv P_{f}^{T,d} - P_{f}^{T,s} + P_{f}^{E,d} - P_{f}^{E,s} ,\qquad 
{\mathcal P}^p_f  \equiv (P_{f}^{T,d} + P_{f}^{T,s} + P_{f}^{E,d} + P_{f}^{E,s} )/2\,.
\end{split}
\eeq
In the latter it is understood that scale and scheme dependence has been subtracted out by $\tilde P_f$.
We can now write the decay amplitudes in terms of the above quantities as 
\beq\label{diagrammaticamps}
\begin{split}
A(\bar D^0\to K^+\pi^-)&=V_{cs} V_{ud}^* \,{\mathcal T}_{K\pi} ,\qquad A(\bar D^0\to \pi^+K^-)=V_{cd} V_{us}^*\, {\mathcal T}_{\pi K}\\
A(\bar D^0\to \pi^+\pi^-)&= -\tfrac{1}{2}\big(V_{cs}V_{us}^*- V_{cd}V_{ud}^*\big)\left(
{\mathcal T}_{\pi\pi}  + {\mathcal P}_{\pi \pi }^t \right) 
- V_{cb}V_{ub}^* \left(\tilde P_{\pi \pi} +  {\mathcal P}_{\pi \pi }^p + {\mathcal T}_{\pi\pi}/2 \right),
\\
A(\bar D^0\to K^+K^-)&= \tfrac{1}{2}\big(V_{cs}V_{us}^*- V_{cd}V_{ud}^*\big)\left(
{\mathcal T}_{KK}  - {\mathcal P}_{K K }^t \right)
- V_{cb}V_{ub}^* \left(\tilde P_{K K} +  {\mathcal P}_{K K }^p + {\mathcal T}_{KK}/2 \right).
\end{split}
\eeq

In the U-spin limit, the non-contracted amplitudes satisfy
\beq  {\mathcal T}_{KK}^s = {\mathcal T}_{\pi\pi}^d = {\mathcal T}_{K\pi} = {\mathcal T}_{\pi K} \equiv T \,, \eeq
where $T$ is defined in \eqref{TandP} -\eqref{eq-t1} in terms of operator matrix elements.
Introducing the U-spin breaking parameters $\epsilon_{T1}$ and $ \epsilon_{T2} $,
we can express these amplitudes at ${\mathcal O}(\epsilon_U)$ as 
\beq
\begin{split} \label{firstorderT}
{\mathcal T}_{KK}^s & = T (1 - \tfrac{1}{2} \epsilon_{T1} ), \qquad {\mathcal T}_{\pi\pi}^d = T (1 + \tfrac{1}{2} \epsilon_{T1} ) \\
{\mathcal T}_{K\pi} & = T (1 - \tfrac{1}{2} \epsilon_{T2} ), \qquad {\mathcal T}_{\pi K} = T (1 + \tfrac{1}{2} \epsilon_{T2} )\,.
\end{split}
\eeq
Similarly, in the U-spin limit, the penguin contractions in the penguin amplitudes satisfy
\beq {\mathcal P}_{K K }^p  =  {\mathcal P}_{\pi \pi }^p \equiv P\,,\eeq
where $P$ is defined in \eqref{eq:P}.   Introducing the U-spin breaking parameter $\epsilon_P$, we can 
write them at ${\mathcal O}(\epsilon_U )$ as 
\beq\label{firstorderP}
{\mathcal P}_{KK}^p  = P (1 + \tfrac{1}{2} \epsilon_{P} ), \qquad {\mathcal P}_{\pi\pi}^p = P (1 - \tfrac{1}{2} \epsilon_{P} ) \,.\eeq
While the ${\mathcal P}^t_f$ vanish in the U-spin limit,
at ${\mathcal O}(\epsilon_U )$ we have
\beq {\mathcal P}^t_{K K } = {\mathcal P}^t_{\pi \pi } \equiv P_{\rm break} \,,\eeq
where $P_{\rm break }$ is defined in terms of operator matrix elements in \eqref{eq:t3}.
$P_{\rm break} $ and $P$ can be related by
\beq \label{firstorderPb} P_{\rm break} = \epsilon_{sd}^{(1)} \, P\,, \eeq
where the U-spin breaking parameter $\epsilon_{sd}^{(1)}$ 
accounts for the difference in $\bar s s$ and $\bar d d$ rescattering contributions at 
${\mathcal O}(\epsilon_U)$.  The difference between ${\mathcal P}^t_{K K }$
and ${\mathcal P}^t_{\pi \pi} $ enters formally at $\epsilon_{U}^2$.  Introducing a 
a new U-spin breaking parameter $\epsilon_{sd}^{(2)}$ to take this difference into account yields the relations
\beq  \label{secondorderPb} {\mathcal P}^t_{K K } = P_{\rm break} (1 + \tfrac{1}{2} \epsilon_{sd}^{(2)})\,,\qquad
{\mathcal P}^t_{\pi \pi } = P_{\rm break} (1 - \tfrac{1}{2} \epsilon_{sd}^{(2)})\,.\eeq

Finally, substituting above expressions \eqref{firstorderT}, \eqref{firstorderP}, \eqref{secondorderPb} for ${\mathcal T}_f $, ${\mathcal P}^p_f $, ${\mathcal  P}^t_f $, respectively, in \eqref{diagrammaticamps} and neglecting the $\tilde P_f$, yields the diagrammatic expressions for the decay amplitudes
given in \eqref{diagrammatic-begin}-\eqref{diagrammatic-end} of Section~\ref{sec:CP}.
We have assumed that the penguin contractions are dynamically enhanced
by ${\mathcal O}(1/\epsilon')$, where $\epsilon' \sim \epsilon_U$ which leads to the scalings for the various amplitude contributions given in 
\eqref{scalings1} and \eqref{scalings2}.



\begin{thebibliography}{99}

\bibitem{Savage:1991wu} 
  M.~J.~Savage, 
 Phys.\ Lett.\ B {\bf 257}, 414 (1991).

\bibitem{Hinchliffe:1995hz} 
  I.~Hinchliffe and T.~A.~Kaeding,
  Phys.\ Rev.\ D {\bf 54}, 914 (1996)
  [hep-ph/9502275].

\bibitem{Ryd:2009uf}
A.~Ryd and A.~A.~Petrov,
Rev.\ Mod.\ Phys.\ {\bf 84} (2012) 65
[arXiv:0910.1265 [hep-ph]].

\bibitem{Cheng:2010ry}
H.~Y.~Cheng and C.~W.~Chiang,
Phys.\ Rev.\ D {\bf 81} (2010) 074021
[arXiv:1001.0987 [hep-ph]].

\bibitem{Pirtskhalava:2011va}
D.~Pirtskhalava and P.~Uttayarat,
arXiv:1112.5451 [hep-ph].

\bibitem{Cheng:2012wr}
H.~Y.~Cheng and C.~W.~Chiang,
arXiv:1201.0785 [hep-ph].

\bibitem{Bhattacharya:2012ah}
B.~Bhattacharya, M.~Gronau and J.~L.~Rosner,
arXiv:1201.2351 [hep-ph].

\bibitem{Feldmann:2012js}
T.~Feldmann, S.~Nandi and A.~Soni,
arXiv:1202.3795 [hep-ph].


\bibitem{Falk:2001hx}
A.~F.~Falk, Y.~Grossman, Z.~Ligeti and A.~A.~Petrov,
Phys.\ Rev.\ D {\bf 65} (2002) 054034
[arXiv:hep-ph/0110317].

\bibitem{Aaij:2011in} 
  R.~Aaij {\it et al.}  [LHCb Collaboration],
  arXiv:1112.0938 [hep-ex].
  
  \bibitem{CDF-talk}
A. Di Canto, talk at 
XXVI Rencontres de Physique de la Vallee d'Aoste Feb 26th-Mar 3rd 2012, La Thuile, Italy;
CDF Note 10784.

\bibitem{Aubert:2007if}
  B.~Aubert {\it et al.}  [BaBar Collaboration],
  Phys.\ Rev.\ Lett.\  {\bf 100}, 061803 (2008)
  [arXiv:0709.2715 [hep-ex]].
  
  \bibitem{Staric:2008rx}
  M.~Staric {\it et al.}  [Belle Collaboration],
  Phys.\ Lett.\  B {\bf 670}, 190 (2008)
  [arXiv:0807.0148 [hep-ex]].
  
  \bibitem{Aubert:2007en} 
  B.~Aubert {\it et al.}  [BABAR Collaboration],
  Phys.\ Rev.\ D {\bf 78}, 011105 (2008)
  [arXiv:0712.2249 [hep-ex]].
  
  \bibitem{Staric:2007dt}
  M.~Staric {\it et al.}  [Belle Collaboration],
  Phys.\ Rev.\ Lett.\  {\bf 98}, 211803 (2007)
  [arXiv:hep-ex/0703036].



%

\bibitem{Grossman:2006jg}
  Y.~Grossman, A.~L.~Kagan, Y.~Nir,
  Phys.\ Rev.\  {\bf D75}, 036008 (2007).
  [hep-ph/0609178].

\bibitem{Isidori:2011qw} 
  G.~Isidori, J.~F.~Kamenik, Z.~Ligeti and G.~Perez,
  arXiv:1111.4987 [hep-ph].

\bibitem{Wang:2011uu}
K.~Wang and G.~Zhu,
arXiv:1111.5196 [hep-ph].

\bibitem{Hochberg:2011ru}
Y.~Hochberg and Y.~Nir,
arXiv:1112.5268 [hep-ph].

\bibitem{Chang:2012gn} 
  X.~Chang, M.~-K.~Du, C.~Liu, J.~-S.~Lu and S.~Yang,
  arXiv:1201.2565 [hep-ph].

\bibitem{Giudice:2012qq} 
  G.~F.~Giudice, G.~Isidori and P.~Paradisi,
  arXiv:1201.6204 [hep-ph].

\bibitem{Altmannshofer:2012ur} 
  W.~Altmannshofer, R.~Primulando, C.~-T.~Yu and F.~Yu,
  arXiv:1202.2866 [hep-ph].

\bibitem{Chen:2012am} 
  C.~-H.~Chen, C.~-Q.~Geng and W.~Wang,
  arXiv:1202.3300 [hep-ph].

\bibitem{Golden:1989qx} 
  M.~Golden and B.~Grinstein,
  Phys.\ Lett.\ B {\bf 222}, 501 (1989).
  
   \bibitem{Buccella:1994nf}
  F.~Buccella, M.~Lusignoli, G.~Miele, A.~Pugliese and P.~Santorelli,
  Phys.\ Rev.\  D {\bf 51} (1995) 3478
  [arXiv:hep-ph/9411286].

\bibitem{Brod:2011re}
J.~Brod, A.~L.~Kagan and J.~Zupan,
arXiv:1111.5000 [hep-ph].  

\bibitem{kaganFPCP} A.~L.~Kagan, talk at FPCP 2011, Kibbutz Maale Hahamisha, Israel, May 2011.

\bibitem{Franco:2012ck}
  E.~Franco, S.~Mishima and L.~Silvestrini,
  arXiv:1203.3131 [hep-ph].
  
  \bibitem{Li:2012cf}
  H.~n.~Li, C.~D.~Lu and F.~S.~Yu,
  arXiv:1203.3120 [hep-ph].
  
\bibitem{Cirigliano:2009rr} 
  V.~Cirigliano, G.~Ecker and A.~Pich,
  Phys.\ Lett.\ B {\bf 679}, 445 (2009)
  [arXiv:0907.1451 [hep-ph]].


\bibitem{Sharpe-private}
S. Sharpe, private communication. 

\bibitem{Asner:2010qj}
D.~Asner {\it et al.} [Heavy Flavor Averaging Group],
arXiv:1010.1589 [hep-ex]; summer 2011 updates available at {\tt http://www.slac.stanford.edu/xorg/hfag/}.


\bibitem{Charles:2011va} 
  J.~Charles, O.~Deschamps, S.~Descotes-Genon, R.~Itoh, H.~Lacker, A.~Menzel, S.~Monteil and V.~Niess {\it et al.},
  Phys.\ Rev.\ D {\bf 84}, 033005 (2011)
  [arXiv:1106.4041 [hep-ph]].
  
  \bibitem{Aaltonen:2011se}
  T.~Aaltonen {\it et al.}  [CDF Collaboration],
  Phys.\ Rev.\  D {\bf 85}, 012009 (2012)
  [arXiv:1111.5023 [hep-ex]].
  


\end{thebibliography}
\end{document}